\documentclass{JHEP3}

\usepackage{enumerate}
\usepackage{amsmath}
\usepackage{amssymb}
\usepackage{amsfonts}
\usepackage{latexsym}
\usepackage{epsfig}
\usepackage{array}
\usepackage{multirow}
\usepackage[vcentermath]{youngtab}

\newcommand\field[1]{{\ensuremath{\mathbb{{#1}}}}}

\newcommand{\BB}{\mathcal{B}}
\newcommand{\RR}{\field{R}}

\newcommand{\sT}{\mathcal{T}} 
\newcommand{\GG}{\mathcal{G}}

\newcommand{\NN}{\mathcal{N}}

\newcommand{\sA}{\mathcal{A}}

\newcommand{\Xh}{{\hat{X}}}
\newcommand{\Sh}{{\hat{S}}}
\newcommand{\Zh}{{\hat{Z}}}

\newcommand{\tk}{{T_\kappa}}
\newcommand{\tn}{{T_\nu}}
\def\Tk#1{{\mathcal{T}_{#1,\kappa}}}
\def\Tn#1{{\mathcal{T}_{#1,\nu}}}

\def\ov{\over}

\def\lam{{\lambda}}

\def\vev#1{\langle#1\rangle}

\def\tr{{\rm tr}}
\def\Tr{{\rm Tr}}
\def\NN{{\cal N}}
\def\SS{{\cal S}}

\def\eq#1{(\ref{#1})}
\def\vev#1{{\langle#1\rangle}}

\def\Om{{\Omega}}

\def \lam {\lambda}
\def \om {\omega}
\def \ra {\rightarrow}

\def\LL{{\cal L}}

\def\n{{V_A}}

\def\LL{{\cal L}}
\def\NN{{\cal N}}
\def\CC{{\cal C}}

\newcommand{\be}{\begin{equation}}
\newcommand{\ee}{\end{equation}}
\newcommand{\bea}{\begin{eqnarray}}
\newcommand{\eea}{\end{eqnarray}}

\newcommand{\bln}{\begin{align}}
\newcommand{\eln}{\end{align}}
\newcommand{\bst}{\begin{split}}
\newcommand{\est}{\end{split}}
\newcommand{\bi}{\begin{itemize}}
\newcommand{\ei}{\end{itemize}}
\newcommand{\ben}{\begin{enumerate}}
\newcommand{\een}{\end{enumerate}}

\title{Anomaly Equations and Intersection Theory}

\author{Daniel S. Park\\
Center for Theoretical Physics\\
Department of Physics\\
Massachusetts Institute of Technology\\
Cambridge, MA 02139, USA\\
\\
\\
{\tt dspark81} {\rm at} {\tt mit.edu}
}

\preprint{MIT-CTP-4315}

\abstract{Six-dimensional supergravity theories with $\NN=(1,0)$
supersymmetry must satisfy anomaly equations. These equations come from
demanding the cancellation of gravitational, gauge and mixed anomalies.
The anomaly equations have implications for
the geometrical data of Calabi-Yau threefolds,
since F-theory compactified on an elliptically fibered
Calabi-Yau threefold with a section generates a
consistent six-dimensional $\NN=(1,0)$ supergravity theory.
In this paper, we show that the anomaly equations can be summarized by
three intersection theory identities.
In the process we also identify the geometric counterpart of
the anomaly coefficients---in particular, those of the abelian
gauge groups---that govern the low-energy dynamics of the theory.
We discuss the results in the context of investigating
string universality in six dimensions.}

\begin{document}

%--------------- ARTICLE ---------------------------
\section{Introduction}

%The chiral fields of theories in
%$(2k+2)$ dimensions are severely restricted due to
%the existence of gravitational, gauge and mixed anomalies
%in these dimensions.
%To be more precise, the effective potential of a
%single chiral field coupled to gravity or gauge fields will
%not be diffeomorphism invariant or gauge invariant.
%The 

The massless spectrum of six-dimensional supergravity theories
with minimal supersymmetry is severely restricted due to
the existence of gravitational, gauge and mixed anomalies
\cite{anomalies,ganomalies}.
This implies that the massless spectrum must satisfy
anomaly equations that come from a generalized version
of the Green-Schwarz factorization condition
\cite{GSW,Sagnotti:1992qw,Sadov},
originally formulated in ten-dimensions \cite{GS}.
These equations involve two types of data of the theory:
\ben
\item The anomaly coefficients, which we schematically denote by $\{b\}$.
\item The spectrum of the theory, which we schematically denote by $S$.
\een
Then the anomaly equations have the form:
\be
f_i (\{b\})=F_i (S)
\ee
where $f_i $ and $F_i$ are some functions.

These equations have implications on the geometry of
elliptically fibered Calabi-Yau threefolds that have a section.
For an elliptically fibered Calabi-Yau threefold $X$ with
a section and a smooth resolution, we can obtain a corresponding
consistent six-dimensional supergravity theory
by compactifying F-theory on $X$
\cite{VafaF,MV1,MV2}.
The resulting low-energy effective theory is guaranteed to
be non-anomalous as the the background is consistent.
Hence if we denote the anomaly coefficients
of this theory $\{b\}(X)$, and the massless spectrum of this theory $S(X)$,
they must satisfy the anomaly equations:
\be
f_i (\{b\}(X))=F_i (S(X))
\ee
Therefore, the anomaly equations imply
non-trivial identities involving the geometric data of
$X$ \cite{AKM,GM2000,GM2011}.\footnote{Currently,
we understand anomaly cancellation
only when the theory obtained by compactifying
F-theory on $X$ has a weakly-coupled field theory description.
When $X$ has singularities whose resolution involves blowing up
a point into a four-cycle \cite{WittenPT}---or equivalently,
when $X$ has codimension-two singularities in the base
whose resolution involves blow-ups of the base
\cite{AM}---one obtains an exotic theory with tensionless strings
\cite{MV2,tensionless1,tensionless1p5,
tensionless2,tensionless3,tensionless4,tensionless5}
by compactifying F-theory on $X$.
We do not consider such $X$, {\it i.e.}, we assume all singularities
of $X$ can be resolved by blowing up rational curves throughout
this paper.}

In order to get to the geometric identities, we must first be able to
identify $\{b\}(X)$ and $S(X)$, given a manifold $X$.
$\{b\}(X)$ are identified readily when all the massless vector fields
of the theory are non-abelian, {\it i.e.,} when for each massless vector field,
there exists another vector field charged under it.
The geometric counterpart of abelian anomaly coefficients---to our
knowledge---was unidentified up to this point.
Our first task is to identify the abelian anomaly coefficients
given $X$.
A convenient method to investigate the abelian sector
of F-theory backgrounds is to use M-theory/F-theory
duality \cite{VafaF}, and take an intersection
theory-based\footnote{Some standard references on intersection theory are
\cite{Fulton} and \cite{GH}.}
approach in the dual M-theory background
\cite{AKM,WittenPT,KMP,FerraraMinasianSagnotti,Grimm,GrimmHayashi}.\footnote{We
are aware of an effort to carefully reconstruct the six-dimensional
effective action based on this approach that is under way \cite{GrimmBonetti}.}

Once $\{b\}(X)$ and $S(X)$ are identified for $X$,
we may translate the anomaly equations to identities of
the geometric data of $X$.
The main result of this paper is that the anomaly equations
imply the following three identities for a smooth Calabi-Yau
manifold $\Xh$ that is a smooth resolution of an elliptically
fibered manifold $X$ with a section:
\[
\fbox{
\addtolength{\linewidth}{-2\fboxsep}%
\addtolength{\linewidth}{-2\fboxrule}%
\begin{minipage}{\linewidth}
\begin{align}
\begin{split}
&\pi(\SS_1 \cdot \SS_2)\cdot \pi(\SS_3 \cdot \SS_4)+
\pi(\SS_1 \cdot \SS_3)\cdot \pi(\SS_2 \cdot \SS_4)+
\pi(\SS_1 \cdot \SS_4)\cdot \pi(\SS_2 \cdot \SS_3) \\
&=\sum_r \prod_{k=1}^4 (c_r \cdot \SS_k)
+ \sum_\rho (2g_\rho-2)\prod_{k=1}^4 (\chi_\rho \cdot \SS_k)
\end{split}
\label{one}
\end{align}
\begin{align}
\begin{split}
6K \cdot \pi(\SS_1 \cdot \SS_2) 
=\sum_r (c_r \cdot \SS_1) (c_r \cdot \SS_2) 
+ \sum_\rho (2g_\rho-2)(\chi_\rho \cdot \SS_1) (\chi_\rho \cdot \SS_2) 
\end{split}
\label{two}
\end{align}
\flushright
when $\SS_k \cdot f =0$ for all $k$.
\begin{align}
30K \cdot K + {1 \ov 2} \chi_\Xh =\sum_r 1
+ \sum_\rho (2g_\rho-2)
\label{three}
\end{align}
\smallskip
\end{minipage}\nonumber
}
\]
Some definitions are needed to explain these identities.
Since $\Xh$ is a smooth resolution of an elliptically fibered
manifold, it must also be a fibered manifold whose
fiber at a generic point of the base
is an elliptic curve.
We define $f$ to be the fiber class
and $\BB$ to be the base.
$\pi : \Xh \rightarrow \BB$ is defined to be
the projection to the base manifold.
$K$ denotes the canonical class of the base manifold.
$c_r$ and $\chi_\rho$ are connected rational curves
that shrink in the fibration limit, {\it i.e.,} as $\Xh \rightarrow X$.
$c_r$ are those that are isolated and $\chi_\rho$ are those
that are fibered over a curve of genus $g_\rho$.
$\chi_\Xh$ in \eq{three} is the Euler characteristic
of $\Xh$.

$\SS_k$ denote four-cycles in the full manifold,
while the dots denote intersection operations.
To be precise:
\ben
\item The dots inside the projection function denote the intersection
products in the full manifold $\hat{X}$. As the intersection is taken between
two four-cycles, the product is a two-cycle in $\Xh$.
\item The dots outside the projection function and on the left hand
side of the equations denote the intersection products in the
base manifold. Since $\SS_i \cdot \SS_j$ is a two-cycle,
its projection $\pi(\SS_i \cdot \SS_j)$ is in general a linear combination
of two-cycles, one-cycles and zero-cycles.
Regarding cycles of dimension smaller than two
as being ``null two-cycles,"
the product, being the intersection of two-cycles in the base,
yields a number as the base manifold is two-complex dimensional.
\item The dots on the right hand side of the equations denote the
intersection products in the full manifold $\hat{X}$. Since the product is taken
between a four-cycle and a two-cycle, the intersection product
is a number.
\een
The claim is that the equations \eq{one} and \eq{two}
hold for any four-cycles $\SS_k$ of the manifold $\hat{X}$
that do not intersect the fiber.

The equations \eq{one}, \eq{two} and \eq{three} follow from
the gauge, mixed and gravitational anomaly cancellation conditions,
respectively. We note that the third equation, coming from gravitational
anomaly cancellation, is implicit in works
from the very early stages of F-theory and has been presented in
various forms in the literature \cite{MV1,MV2,AKM,GM2000,GM2011}.

The structure of this paper is as follows.
In section \ref{s:6D} we review six-dimensional theories
with $\NN =(1,0)$ supersymmetry and the anomaly
cancellation conditions.
We define anomaly coefficients and show how they
determine the low-energy dynamics of the theory in this section.
In section \ref{s:6DFthy}, we study F-theory compactified
on Calabi-Yau threefolds.
We use the duality between M-theory and F-theory to
extract the anomaly coefficients and the massless spectrum
from the geometry. The main result of this section is that
we identify the geometric counterpart of abelian anomaly coefficients.
We also show \eq{three} in this section.
In section \ref{s:intersection}, we explain how the anomaly equations
imply the geometric identities \eq{one} and \eq{two}.
Finally in section \ref{s:discussion}, we summarize
the results and discuss its implications.
In particular, we focus on explaining why identifying
the abelian anomaly coefficients in F-theory is an important step
in investigating the question of string universality in six dimensions
\cite{SUniversality}.

\section{6D $(1,0)$ Theories and Anomaly Cancellation}  \label{s:6D}

In this section we review six-dimensional theories
with $\NN=(1,0)$ supersymmetry and the dynamics of
the massless spectrum.
In section \ref{ss:6D} we present an overview
of the field content of six-dimensional theories with
$(1,0)$ supersymmetry.
Next we review anomaly cancellation conditions
for these theories in section \ref{ss:factcond}.
In particular, we introduce the definition of anomaly coefficients
and present how they determine the dynamics of the massless fields.
We summarize in section \ref{ss:6Dsummary}.\footnote{A more
detailed treatment of the contents of this section can be found in \cite{PT}.}

\subsection{The Massless Spectrum} \label{ss:6D}

The massless spectrum of the models we consider
contains four different multiplets of the supersymmetry algebra:
the gravity and tensor multiplet,
vector multiplet, and hypermultiplet. The contents of
these multiplets are summarized in table \ref{t:mult}.

We consider theories with one gravity multiplet.
There can in general be multiple tensor multiplets, whose number
we denote by $T$.
Theories with $T$ tensor multiplets have a moduli space with
$SO(1,T)$ symmetry; the $T$ scalars in each multiplet
combine into a $SO(1,T)$ vector $j$ which can be taken to have
unit norm. The theory may have an arbitrary gauge group.

We write the gauge group for a given theory as
\be
\GG = \prod_{\kappa=1}^{N} \GG_\kappa \times \prod_{i=1}^{\n} U(1)_i \,,
\ee
modulo possible quotients of discrete subgroups.
Since we discuss only local gauge anomalies in this paper,
such quotients can be ignored for our purposes.
The lowercase greek letters $\kappa,\lambda,\cdots$ are
used to denote the simple non-abelian gauge group factors;
lowercase roman letters $i,j,k,\cdots$ are used to denote $U(1)$
factors. $N$ and $V_A$ are used to denote the number of
non-abelian and abelian gauge group factors of the theory.

As explained in \cite{PT}, abelian vector multiplets can become massive
at the linear level by the St\"uckelberg mechanism and form a long
multiplet. We have argued there that such long multiplets can be
safely ignored when discussing the massless spectrum.

The theory also may have hypermultiplets transforming under various
representations of the gauge group. The type of allowed matter
is determined by anomaly cancellation conditions, as we review shortly.

\begin{table}[t!]
\center
  \begin{tabular}{ | c | c |}
  \hline
  Multiplet & Field Content\\ \hline
  Gravity & $(g_{\mu \nu}, \psi^+_\mu, B^+_{\mu \nu}) $    \\  \hline
  Tensor & $(\phi, \chi^-, B^-_{\mu \nu})$     \\  \hline
  Vector & $(A_{\mu}, \lam^+)$     \\  \hline
  Hyper & $(4\varphi, \psi^-)$    \\  \hline
    \end{tabular}
  \caption{Six-dimensional (1,0) supersymmetry multiplets. The signs on
   the fermions indicate the chirality. The signs on antisymmetric tensors
   indicate self-duality/anti-self-duality.}
\label{t:mult}
\end{table}

\subsection{The Anomaly Equations and Anomaly Coefficients} \label{ss:factcond}

In six-dimensional chiral theories there can be gravitational, gauge and
mixed anomalies \cite{anomalies,ganomalies}. The sign with which each
chiral field contributes to the anomaly is determined
by their chirality.
The 6D anomaly can be described by the method of descent from
an 8D anomaly polynomial. The anomaly polynomial
is obtained by adding up the contributions of all the
chiral fields present in the theory.
The anomaly polynomial is given by \cite{PT};
\begin{align}
\label{raw}
\begin{split}
I_8=
&-{1 \ov 5760} (H-V+29T-273) [\tr R^4+{5 \ov 4} (\tr R^2)^2] \\
&-{1\ov 128} (9-T) (\tr R^2)^2 \\
&-{1 \ov 96} \tr R^2 [\sum_\kappa \Tr F_\kappa^2
 -\sum_{\kappa,R} x_R \tr_{R}F_\kappa^2] \\
&+{1\ov 24} [\sum_\kappa \Tr F_\kappa^4
 -\sum_{R,\kappa} x_R \tr_{R}F_\kappa^4
 -6\sum_{\kappa,R,\lambda,S} x_{RS}
 (\tr_{R}F_\kappa^2)(\tr_{S}F_\mu^2)] \\
 &+{1 \ov 96} \tr R^2 \sum_{i,j,q_i,q_j} x_{q_i q_j} q_{i}q_{j} F_i F_j  \\
&-{1 \ov 6} \sum_{\kappa,R,i,q_i} x_{R,q_i} q_{i}
 (\tr_{R}F_\kappa^3) F_i 
-{1 \ov 4} \sum_{\kappa,R,i, j,q_i,q_j} x_{R,q_i,q_j} q_{i} q_{j}
 (\tr_{R}F_\kappa^2) F_i F_j \\
&-{1 \ov 24}\sum_{i,j,k,l,q_i,q_j,q_k,q_l} x_{q_i,q_j,q_k,q_l} q_{i} q_{j} q_{k}q_{l} F_i F_j F_k F_l
\end{split}
\end{align}
$V$ and $H$ are the numbers of
massless vector multiplets and hypermultiplets in the theory.
We use ``$\tr$" to denote the trace in the fundamental representation,
and ``$\Tr$" to denote the trace in the adjoint. $\tr_R$ denotes the trace in
representation $R$.
The various $x$'s denote the number of hypermultiplets of given
charge or representation:
\ben
\item $x_R$ is the number of hypermultiplets of representation
$R$ of gauge group $\GG_\kappa$.
\item $x_{RS}$ is the number of hypermultiplets of representation
$R \times S$ of gauge group $\GG_\kappa \times \GG_\mu$.
\item $x_{R,q_i}$ is the number of hypermultiplets of representation
$R$ of gauge group $\GG_\kappa $ with charge $q_i$ under $U(1)_i$.
\item $x_{R,q_i,q_j}$ is the number of hypermultiplets of representation
$R$ of gauge group $\GG_\kappa$ with charge $(q_i, q_j)$ under $U(1)_i \times U(1)_j$.
\item $x_{q_i,q_j,q_k,q_l}$ is the number of hypermultiplets
that have charge $(q_i, q_j, q_k, q_l)$ under $U(1)_i \times U(1)_j\times U(1)_k\times U(1)_l$.
\een
Multiplication of forms should be interpreted
as wedge products throughout this paper unless stated otherwise.

In the presence of multiple tensors in the theory, a generalized
version \cite{GSW,Sagnotti:1992qw,Sadov}
of the Green-Schwarz mechanism \cite{GS} can be
used to cancel the anomaly when the anomaly polynomial
can be ``factorized" into a certain form.
To be precise, when there are $T$ tensor multiplets in the theory,
the anomaly polynomial must be ``factorized" in the following way:
\be
\label{genfactorization0}
I_8=-{1 \ov 32} \Om_{\alpha \beta} X_4^\alpha X_4^\beta
\ee
Here $\Om$ is a symmetric bilinear form(or metric) in
$SO(1,T)$ and $X_4$ is a four-form that is an $SO(1,T)$ vector.
$X_4$ can be written as
\be
X_4^\alpha = {1 \ov 2} a^\alpha \tr R^2
+ \sum_\kappa ({2b_\kappa^\alpha \ov \lambda_\kappa})
\tr F_\kappa^2 + \sum_{ij} 2 b^\alpha_{ij} F_i F_j
\label{x4}
\ee
where we define $b_{ij}$ to be symmetric in $i,j$. The $a$
and $b$'s are $SO(1,T)$ vectors and $\alpha$ are $SO(1,T)$ indices.
$a$, $b_\kappa$, and $b_{ij}$ are the anomaly coefficients of the theory.
 
Note that in the presence of $U(1)$'s, the anomaly polynomial factorization
condition is generalized in the form \eq{x4} due to the fact that the field
strength is gauge invariant on its own \cite{Riccioni}.
Under linear redefinitions of the $U(1)$'s $b_{ij}$ transforms as
a bilinear, {\it i.e.,}
\be
F_i= M_i^j F_j', \quad b_{ij} ' = M_i^k M_j^l b_{kl},
\quad M \in GL(V_A,\RR) \,.
\ee

The $\lambda_\kappa$'s are normalization factors that are fixed
by demanding that the smallest topological charge of an embedded
$SU(2)$ instanton is 1. $\lambda_\kappa$ is actually equal to
the Dynkin index of the fundamental representation of the gauge group $\GG_\kappa$.
The values of $\lambda_\kappa$ for given $\GG_\kappa$ are listed in table \ref{t:lambda}
for all the simple groups. $b_\kappa$ turn out to form an integral
$SO(1,T)$ lattice when we include these normalization
factors \cite{KMT2}.

\begin{table}[t!]
\center
  \begin{tabular}{ | c | c | c | c | c | c | c | c | c | c |}
  \hline
 $\GG$ & $A_n$ & $B_n$ & $C_n$ & $D_n$ & $E_6$ &$E_7$ & $E_8$ & $F_4$ & $G_2$ \\ \hline
 $\lambda$ & 1 &2& 1& 2& 6 &12& 60 &6 &2 \\ \hline
    \end{tabular}
  \caption{Normalization factors for the simple groups.}
\label{t:lambda}
\end{table}

There is an important fact related to the factors $\lambda(\GG)$
worth noting for future reference.
Let us define the normalized basis $\{ T_i \}$
for the Cartan sub-algebra of $\GG$ such that
\be
\tr_f T_i T_j = \delta_{ij}\,.
\ee
This provides an unambiguous normalization for the root lattice
of a given Lie group.
Note that two Lie groups with the same Lie algebra can have different
normalizations of the root lattice if their fundamental representations differ.
We may define the ``coroot basis" for the Cartan sub-algebra as
\be
\sT_I \equiv {2\alpha_I^i T_i \ov \langle \alpha_I, \alpha_I \rangle} 
\ee
where $\alpha^i_I$ are the coordinates of the $I$'th simple root.
$\sT_I$ have the following properties:
\ben
\item The charge of the root vector $E_{\alpha}$ under
$\sT_I $ is
\be
\sT_I | \alpha \rangle
= {2\langle \alpha_I, \alpha \rangle \ov \langle \alpha_I, \alpha_I \rangle}   | \alpha \rangle \,.
\ee
In particular for the simple roots of the Lie algebra,
\be
\sT_I | \alpha_J \rangle
= C_{IJ}  | \alpha_J \rangle \,.
\ee
where $C_{IJ}$ are the elements of the Cartan matrix.
We note that the Cartan matrix is determined
by the gauge algebra, rather than the gauge group.
For example, it is the same for $SO(3)$ and $SU(2)$.
\item The charge of any weight vector $| \beta \rangle$ under
$\sT_I $ is
\be
{2\langle \alpha_I, \beta \rangle \ov \langle \alpha_I, \alpha_I \rangle}  \,,
\ee
which is always integral, by definition of weight vectors.
\item For two basis elements among $\{ \sT_I \}$,
\be
{1 \ov \lambda(\GG)}\tr \sT_I \sT_J =  \CC_{IJ}
\ee
where $\CC$ is the normalized inner product matrix of the coroots, {\it i.e.,}
\be
\CC_{IJ} =  {1 \ov \lambda(\GG)}{4  \langle \alpha_I, \alpha_J \rangle \ov
 \langle \alpha_I, \alpha_I \rangle \langle \alpha_J, \alpha_J \rangle} \,.
\ee
Just as with the Cartan matrix, the normalized coroot matrix $\CC$
is determined uniquely by the gauge algebra.
\een
Proofs of these statements and the relation of the normalized
coroot matrix $\CC$ to intersection theory is discussed in
appendix \ref{ap:lieint}.

The gauge invariant three-form field strengths are given by
\be
H^\alpha=dB^\alpha+{1 \ov 2} a^\alpha \om_{3L}
+2\sum_\kappa {b^\alpha_\kappa \ov \lambda_\kappa} \om^\kappa_{3Y}
+2\sum_{ij} {b^\alpha_{ij}} \om^{ij}_{3Y} \,.
\label{fstrength}
\ee
$\om_{3L}$ and
$\om_{3Y}$ are Chern-Simons 3-forms of the spin connection
and gauge field respectively. If the factorization condition
(\ref{genfactorization0}) is satisfied, anomaly cancellation can
be achieved by adding the local counter-term
\be
\delta \LL_{GS} \propto -\Om_{\alpha\beta} B^\alpha \wedge X_4^\beta \,.
\label{GSterm}
\ee

Supersymmetry determines the kinetic term for the gauge
fields to be (up to an overall factor) \cite{Sagnotti:1992qw,Riccioni}
\be
-\sum_\kappa ({j \cdot b_\kappa \ov \lambda_\kappa}) \tr F_\kappa^2
-\sum_{k,l} (j \cdot b_{kl}) F_i F_j \,,
\label{kinetic}
\ee
where the squared terms of the field strength are inner products.
$j$ is the unit $SO(1,T)$ vector defined by the $T$ scalars in the tensor
multiplets. The inner product of $j$ and the $b$ vectors are defined
with respect to the metric $\Om$. We want there to be a value of
$j$ such that all the gauge fields have positive definite kinetic terms.
This means that there should be some value of $j$ such that
all $j \cdot b_\kappa$ are positive and that $j \cdot b_{kl}$
is a positive definite matrix with respect to $k,l$.

In the presence of abelian vector multiplets,
there is yet another way to cancel anomalies \cite{Berkooz:1996iz},
by coupling the abelian vector field to a St\"uckelberg
zero-form field. It is shown in \cite{PT} that this mechanism can be
safely ignored when we are discussing the massless vector fields.
In other words, it is shown that the mixed/gauge anomalies
of the unbroken gauge group are cancelled by the tensor fields,
not by zero-form fields.

The anomaly equations come from demanding that the anomaly
polynomial \eq{raw} coming from adding all the contributions from the
chiral fields in the theory factorize in the form
\eq{genfactorization0}.
This amounts to the following equations.
\\
\medskip

\noindent\textbf{Equations from gravitational anomalies}
\[
\fbox{
\addtolength{\linewidth}{-2\fboxsep}%
\addtolength{\linewidth}{-2\fboxrule}%
\begin{minipage}{\linewidth}
\begin{align}
\begin{split}
273&=H-V+29T \\
a\cdot a&= 9-T 
\end{split}
\label{grav}
\end{align}
\end{minipage}\nonumber
}
\]
These equations come from demanding that
pure gravitational anomalies are cancelled.
Here, $H$ denotes the number of hypermultiplets
and $V$ denotes the number of vector multiplets.
%Since uncharged hypermultiplets do not show up in any of the equations
%other than here, this implies that the number of charged hypermultiplets are
%bounded by $V+273-29T$.
%As long as the number of charged hypermultiplets is below this bound,
%we can add neutral hypermultiplets to the theory and fill up the discrepancy.
\\
\medskip

\noindent\textbf{Equations from mixed anomalies}
\[
\fbox{
\addtolength{\linewidth}{-2\fboxsep}%
\addtolength{\linewidth}{-2\fboxrule}%
\begin{minipage}{\linewidth}
\begin{align}
\begin{split}
a\cdot ( {b_\kappa \ov \lambda_\kappa} )&=
{1 \ov 6} (A_{\text{Adj}_\kappa}-\sum_R x_R A_R)\\
a \cdot b_{ij} &= -{1 \ov 6} \sum_I x_{q_i,q_j} q_{i} q_{j}
\end{split}
\label{mixed}
\end{align}
\end{minipage}\nonumber
}
\]
These equations should be satisfied for each gauge group $\GG_\kappa$,
$U(1)_i$, and $U(1)_j$.
The inner products on the left-hand-side of the equations
are inner products with respect to the $SO(1,T)$ metric $\Om$.
The group theory factor $A_R$ is defined to be
\be
\tr_R F_\kappa^2 = A_R \tr F_\kappa^2
\ee
for a given representation $R$ of the gauge group $\GG_\kappa$.
\\
\medskip

\noindent\textbf{Equations from gauge anomalies}
\[
\fbox{
\addtolength{\linewidth}{-2\fboxsep}%
\addtolength{\linewidth}{-2\fboxrule}%
\begin{minipage}{\linewidth}
\begin{align}
\begin{split}
0&= B_{\text{Adj}_\kappa}-\sum_R x_R B_R \\
({b_\kappa \ov \lambda_\kappa})^2&=
{1 \ov 3}  (\sum_R x_R C_R-C_{\text{Adj}_\kappa})\\
({b_\kappa \ov \lambda_\kappa}) \cdot ({b_\mu \ov \lambda_\mu})&=
\sum_I x_{RS} A_R A_S \\
0 &= \sum_{R,q_i} x_{R,q_i}  q_{i}  E_R \\
({b_\kappa \ov \lambda_\kappa}) \cdot b_{ij} &=
\sum_{R,q_i,q_j} x_{R,q_i,q_j} q_{i} q_{j} A_R \\
b_{ij} \cdot b_{kl} + b_{ik} \cdot b_{jl} + b_{il} \cdot b_{jk}
&=  \sum_{q_i,q_j,q_k,q_l} x_{q_i,q_j,q_k,q_l} q_{i} q_{j} q_{k} q_{l} 
\end{split}
\label{gauge}
\end{align}
\end{minipage}\nonumber
}
\]
These equations should be satisfied for all $\GG_\kappa \neq \GG_\lambda$, and
for all $U(1)_i$, $U(1)_j$, $U(1)_k$ and $U(1)_l$.
For each representation $R$ of
group $\GG_\kappa$ the group theory coefficients
$B_R$ and $C_R$ are defined by
\be
\tr_R F^4 = B_R \tr F^4 + C_R (\tr F^2)^2
\ee
In the event that there is only one fourth order invariant for
the given gauge group---as is with for example, $SU(2)$---we
define $B_R=0$.
Also, $E$ is defined by
\be
\tr_R F^3 = E_R \tr F^3 \,.
\ee

It was shown in \cite{KMT2} using the anomaly equations that
the $SO(1,T)$ vector $a$
and the non-abelian anomaly coefficients $b_\kappa$
form an integral lattice $\Lambda$. It was subsequently shown
that quantum consistency conditions impose that $\Lambda$ must
further be embeddable in a unimodular lattice \cite{SeibergTaylor}.

\subsection{Summary} \label{ss:6Dsummary}

A six-dimensional $\NN=(1,0)$ theory is characterized by its
massless spectrum, the vacuum expectation value of the
scalars present in the theory, and the anomaly coefficients
$a, b_\kappa$ and $b_{ij}$.
The massless spectrum is specified by the following data:
\ben
\item The number of tensor multiplets $T$.
\item The gauge group
\be
\GG = \prod_{\kappa=1}^{N} \GG_\kappa \times \prod_{i=1}^{\n} U(1)_i \,.
\ee
\item The hypermultiplet matter content.
\een
The vacuum expectation value of the scalars in the tensor multiplet
is given by a $SO(1,T)$ unit vector $j$.
The anomaly coefficients are $SO(1,T)$ vectors,
and in particular $b_{ij}$ is also a bilinear form which transforms
under the linear redefinitions of the $U(1)$'s.
The massless matter content and the anomaly coefficients must
satisfy the anomaly equations \eq{grav}, \eq{mixed} and \eq{gauge}.

The anomaly coefficients determine the invariant field strength of
the tensors \eq{fstrength}, the Green-Schwarz term of the quantum effective
action \eq{GSterm}, and the corrected kinetic term for the gauge fields \eq{kinetic}.
Quantum consistency conditions demand that $a, b_\kappa$ must be
embeddable into a unimodular lattice.

\section{Six-dimensional F-theory Backgrounds} \label{s:6DFthy}

F-theory is a convenient way of thinking about
type IIB backgrounds with a varying axio-dilaton \cite{VafaF}.
F-theory backgrounds can be thought of as
being obtained from a twelve-dimensional theory
compactified on an elliptically fibered manifold with a section.
F-theory, when compactified on a Calabi-Yau threefold,
yields a six-dimensional theory with $\NN=(1,0)$ supersymmetry
\cite{VafaF,MV1,MV2}.

There is an elegant picture of how the geometry of the
internal Calabi-Yau threefold is encoded in the non-abelian sector
of the low-energy theory \cite{GM2000,GM2011,KMT2,KMT1}.
In particular, there is a precise geometric meaning of the non-abelian
anomaly coefficients of the six-dimensional theory.
In this section, we clarify the geometric data
encoded in the abelian sector of the low energy theory,
{\it i.e.,} understand what the abelian anomaly coefficients mean geometrically.

In section \ref{ss:revKMT} we review how the geometry
of the Calabi-Yau threefold is encoded in the non-abelian sector of
the low-energy theory following \cite{KMT2,KMT1}.
In section \ref{ss:Mthy} we use the duality between
M-theory and F-theory to reaffirm the results on the non-abelian sector,
and identify the geometric quantity corresponding to the
abelian anomaly coefficient.
In the process we identify the M-theory origin of the various
fields present in the six-dimensional theory.
We summarize the results in section \ref{ss:summ}.

While our main purpose for
using M-theory/F-theory duality is to
identify the anomaly coefficients
of the abelian gauge fields,
one can also use the same tools to
reconstruct the effective action of the
six-dimensional theory in more detail.
We are aware of the work \cite{GrimmBonetti},
where the authors carry out this analysis
carefully in the case that the gauge group is non-abelian.

\subsection{A Very Brief Review of the Non-Abelian Sector} \label{ss:revKMT}

F-theory backgrounds can be thought of as
non-perturbative type-IIB backgrounds with
seven-branes which are not necessarily mutually local.
When we are compactifying F-theory on some elliptically
fibered manifold,
the base of the manifold $\BB$ can be thought of as the
space we are compactifying type IIB string theory on.
The value of the axio-dilaton varies over the base;
this value is identified with the
complex structure of the elliptic fiber of the fibration.
In order to get $\NN=(1,0)$ supersymmetry,
the total space of the fibration must be a Calabi-Yau threefold.
This fact places restrictions on $\BB$ \cite{KMT2}.
Two relevant conditions that $\BB$ must satisfy are that
$h^{2,0}(\BB)=h^{1,0}(\BB)=0$
and that $K \cdot K = 10-h^{1,1}(\BB)$ where $K$
is the canonical class of $\BB$.
We note that the first condition implies that
$H_2 (\BB)\cong H^{1,1}(\BB)$.

The fiber degenerates on complex codimension-one
submanifolds of the base. These submanifolds can be thought of as the
submanifolds the seven-branes wrap. The type of degeneration
determines the nature of the seven-brane and tells us 
the non-abelian gauge group we get in the six-dimensional theory
\cite{MV1,MV2,AKM,GM2000,GM2011,BIKMSV}.
The codimension-two singularities can be thought
of as intersecting points of the seven-branes.
These contain information on the local matter we obtain
\cite{GM2000,GM2011,BIKMSV,KV,MTMatter,EsoleYau,EFY,MS,
KrauseMayrhoferWeigand}.

A beautiful fact is that the geometrical data of
an elliptically fibered Calabi-Yau threefold can be encoded in
an integral lattice.
More precisely, geometric data such as the canonical divisor class,
the K\"ahler class of the base manifold, or the algebraic two-cycles
(or divisors) the seven-branes wrap can be expressed as
a vector in the $H_2 \cong H^{1,1} $ lattice of the base manifold.
It turns out that this integral lattice
is precisely the lattice $\Lambda$ that parametrizes the low energy theory.

To provide the complete map between geometrical data and
the low-energy data, we begin by noting that the number of tensor multiplets $T$
satisfies $T = h^{1,1}(\BB)-1$.
The $H_2$ lattice of the base manifold is an $SO(1,T)$ lattice \cite{KMT2}.
The anomaly coefficients and the modulus $j$---which parameterizes
the vacuum expectation values of the scalars in the tensor multiplets---are
vectors in this lattice, and hence correspond to two-cycles in the base.
The $a$ vector corresponds to the canonical divisor class of the base,
the $j$ vector corresponds to the K\"ahler class of the base,
and the $b_\kappa$ vectors corresponds to the locus of brane $\kappa$.
The ``type of degeneration" along the locus $b_\kappa$---more precisely
the singularity type and the monodromy of the fiber---determines the
gauge group $\GG_\kappa$.\footnote{More precisely, the singularity
determines the gauge algebra rather than the gauge group.
This distinction can be ignored for our purposes.}

\subsection{M/F-theory Duality and Abelian Anomaly Coefficients} \label{ss:Mthy}

As appealing the picture drawn so far may be,
it is not clear how to probe the abelian sector of
F-theory backgrounds directly.
This is because the degrees of freedom of the
underlying twelve-dimensional
theory and their interactions---if they exist at all---is
unclear at the moment.
We did not necessarily need such a global picture
to probe the non-abelian sector of the theory as its
dynamics could be determined locally---only
the near-brane geometry mattered in understanding it.
However, there is a global flavor to
abelian gauge symmetry.\footnote{Determining
the abelian gauge symmetry
is subtle in a wide variety of string constructions,
including heterotic, orbifold, intersecting brane, 
fractional brane and F-theory models
\cite{GSW,MV1,MV2,Berkooz:1996iz,Strominger,Erler:1993zy,
Aldazabal,Berglund,Aspinwall,Blumenhagen,Honecker,
Buican:2006sn,Hayashi,GrimmWeigand,MSS,Marsano,DMSS,GKPW}.
The subtlety comes from the fact that abelian vector fields
in the spectrum that are naively massless
can be lifted at the linear level by coupling to
St\"uckelberg fields.}
For example, even simple information such as the number of
massless abelian vector fields is encoded in global data of the
full manifold \cite{MV2,Wazir,HulekKloosterman}.
In order to understand the abelian sector of the theory,
it turns out to be more convenient to take
an intersection theory-based approach
in the M-theory dual of the F-theory background
\cite{AKM,WittenPT,KMP,GrimmHayashi}.
By recovering the low-energy data of
the six-dimensional F-theory background using this duality,
we can identify the geometric
meaning of the abelian anomaly coefficients
by comparing the coefficients of topological terms
obtained from both sides
\cite{FerraraMinasianSagnotti,Grimm,GrimmHayashi}.

We proceed in three steps.
In section \ref{sss:mf}, we first review
M-theory/F-theory duality
and obtain basic information about the massless spectrum of
F-theory from the M-theory side.
In the process we obtain \eq{three}.
In section \ref{sss:nonab}, we demonstrate
how M-theory/F-theory duality can be used to recover
the low-energy data of the non-abelian sector.
Finally in section \ref{sss:ab}, we identify 
the geometric counterparts
of the low-energy data of the abelian sector
in an analogous way.

\subsubsection{M/F-theory Duality} \label{sss:mf}

The duality between M and F-theory \cite{VafaF}
provides the clearest way to see how the low-energy dynamics of
gauge bosons and matter content arise in
F-theory backgrounds.\footnote{A great
review on F-theory and M-theory/F-theory
duality can be found in \cite{Denef}.}
F-theory compactified on $X\times S^1$---where $X$
is an elliptically fibered Calabi-Yau threefold with a section---is
dual to M-theory compactified on $X$.
In the five-dimensional M-theory background, all the K\"ahler deformations
of $X$ become available, unlike in the six-dimensional theory.
These moduli on the F-theory side are given by the size of the $S^1$
and the Wilson lines of the gauge fields along the $S^1$.
By turning these moduli on to generic values,
we may resolve the singular manifold $X$ to $\Xh$.
This is equivalent to going to the Coulomb branch of
the non-abelian gauge theory,
as the five-dimensional vector multiplet has a real adjoint scalar.
We can recover the fibration limit $\Xh \ra X$ as
we turn off all the Wilson lines and take the radius of the $S^1$
to infinity.
In this sense, the six-dimensional theory can be thought of as
a ``decompactification limit" of the M-theory background.
We use the terms ``decompactification limit," ``F-theory limit,"
and ``fibration limit" interchangeably.

Now let us recover the massless spectrum of the six-dimensional
theory from the geometrical data of $\Xh$.
When we compactify the  six-dimensional theory with $\NN=(1,0)$ supersymmetry
on $S^1$, we get a five-dimensional $\NN=2$ theory with 8 supercharges.
The short multiplets of the six-dimensional theory descend to
short multiplets of the five-dimensional theory as shown in table \ref{t:KK}.
By resolving $X$ to $\Xh$ we have turned on Wilson lines,
and hence all multiplets charged under the Cartan sub-algebra
of the full gauge group become massive.
We will denote these multiplets ``charged multiplets."
Charged multiplets descend from either
vector multiplets or hypermultiplets.
Therefore the six-dimensional massless spectrum can be
recovered from the five-dimensional theory
by identifying the massless multiplets and the charged multiplets
that become massless in the decompactification limit.

There is nothing special about the Cartan basis. The Wilson
lines turned on are generic and mutually commuting, and hence
we can always find a Cartan subgroup of which the Wilson lines
are elements of.
We note that the Cartan sub-algebra of the full gauge algebra
consists of the direct sum of the Cartan sub-algebra of the individual
gauge groups. For abelian groups, the Cartan sub-algebra is equal
to the full gauge algebra.

\begin{table}[t!]
\center
  \begin{tabular}{ | c | c |}
  \hline
  6D & 5D\\ \hline
  Gravity & 1$\times$(Gravity)+1$\times$(Vector)   \\  \hline
  Tensor & 1$\times$(Vector)  \\  \hline
  Vector &  1$\times$(Vector)  \\  \hline
  Hyper &   1$\times$(Hyper)  \\  \hline
    \end{tabular}
  \caption{Six-dimensional (1,0) supersymmetry multiplets
  and their descendants in five dimensions when compactified on a circle.}
\label{t:KK}
\end{table}

Let us first identify the massless fields of the five-dimensional theory
\cite{Cadavid,Antoniadis,FerraraKhuriMinasian}.
M-theory compactified on the fully resolved manifold $\Xh$
has $h^{2}(\Xh) =h^{1,1}(\Xh)$ massless vector fields coming from descending
the three-form on the harmonic two-forms of $\Xh$.
Among these, one vector field is inside the
five-dimensional gravity multiplet
and the others belong to vector multiplets.
The two-forms are Poincar\'e dual to four-cycles in $\Xh$, that is,
for any harmonic two-form $\om$ there exists a four-cycle
$\Sigma$ in $\Xh$
such that for any two-cycle $c$ in $\Xh$,
\be
\int_c \om = \Sigma \cdot c
\ee
where the right-hand side denotes the intersection number
between the two cycles.
Therefore, for each massless vector field, there is
a corresponding four-cycle.
On the F-theory side,
one of these vector fields come from KK-reducing the graviton
along the $S^1$, while $(T+1)=h^{1,1}(\BB)$ come from KK-reducing
the one self-dual and $T$ anti-self dual tensor fields.
The rest come from vectors in the six-dimensional vector multiplets
that are either abelian or in the Cartan of a non-abelian gauge group.

Also, there are $h^{3}(\Xh) =h^{2,1}(\Xh)+1$
massless hypermultiplets in the five-dimensional spectrum.
In the decompactification limit, all of these hypermultiplets
become six-dimensional neutral hypermultiplets---hypermultiplets
that are not charged under any vector field in the Cartan.

Now let us identify the charged multiplets.
These come from M2 branes wrapping complex
curves of $\Xh$. Since the charged multiplets should become
massless in the decompactification limit,
they should come from M2 branes wrapping curves
that shrink in the fibration limit.
As we move along the Coulomb branch
to recover the full non-abelian
gauge symmetry of $X$,
two types of curves shrink to zero size.
\begin{enumerate}
\item \textbf{Type I : }Isolated rational curves
that shrink to zero size in the limit $\Xh \rightarrow X$.
\item \textbf{Type F : }Rational curves fibered over a curve
that shrink to zero size in the limit $\Xh \rightarrow X$.
\end{enumerate}
These curves are all rational curves; they are topologically
$\field{P}^1$'s as only these types of
curves can shrink in $\Xh$ \cite{AspinwallK3}.
We index the curves of type I by $r$ and denote them $c_r$,
and index the curves of type F by $\rho$ and denote them $\chi_\rho$.
We use $g_\rho$ to denote the genus of the curve
$\chi_\rho$ is fibered over.
The curve a type F curve is fibered over is either a
curve in the base or its branched cover \cite{AKM}.
In the fibration limit $\Xh \ra X$, the type F curves shrink into
points on the singular fibers along codimension-one loci
in the base while the type I curves shrink into points on
singular fibers at codimension-two loci.

A curve of type I contributes one hypermultiplet,
while a curve of type F fibered over a curve of genus $g$
contributes $2g$ hypermultiplets
and $2$ vector multiplets to the BPS spectrum \cite{WittenPT,KMP}.
By quantizing the zero mode of
an M2 brane wrapping a curve of type I,
one obtains a half-hypermultiplet.
Together with another half-hypermultiplet that comes
from quantizing the zero modes of an anti-brane wrapping the same curve,
a curve of type I contributes one hypermultiplet.
Meanwhile, $2g$ half-hypermultiplets
and one vector multiplet come from quantizing
the zero-modes of an M2 brane wrapping a curve
of type F fibered over a curve of genus $g$.
Also the same number
of multiplets come from quantizing the zero-modes
of an anti-M2 brane wrapping the same curve.
By definition, these multiplets become massless in the
decompactification limit, and are in the massless spectrum of the
six-dimensional theory.

The charge of a charged BPS particle
coming from a brane wrapping a rational curve $c$
under a vector field $A_\Sigma$
coming from reducing the eleven-dimensional three-form field on the
harmonic three-form $\om$ is given by
\be
\pm \int_c \om = \pm \Sigma \cdot c \,,
\ee
where $\Sigma$ is the four-cycle that is Poincar\'e dual to $\om$.
The sign depends on whether the brane is an M2 brane
or and anti-M2 brane.
While the charge of a vector multiplet is unambiguous,
there is an overall sign ambiguity in defining charges
of the hypermultiplet.
A hypermultiplet consists of
two half-hypermultiplets with opposite charges
under any abelian or non-abelian Cartan gauge field;
one coming from M2 branes wrapped on a curve
and the other coming from an anti-brane wrapped
on the same curve.
We fix the sign of the charge
of a hypermultiplet coming from a curve $c$
under a gauge field $A_\Sigma$ as
\be
\int_c \om = \Sigma \cdot c \,.
\ee

Meanwhile, the charge of vector or hypermultiplets
under the vector multiplets
coming from shrinking rational curves of type F can be
obtained by considering the algebra of BPS states in the
Calabi-Yau manifold as described in \cite{HM}.
Some of the multiplets that are uncharged under the
vector fields in the Cartan sub-algebra
can be charged under vector fields that come from shrinking type F
rational curves.

Let us summarize what we have learned.
There are $h^{1,1}(\Xh)$ massless vector fields in the five-dimensional
M-theory background. In the decompactification limit,
two of them belong to the gravity multiplet, $T=(h^{1,1}(\BB)-1)$ of them
belong to the tensor multiplets and the rest of them belong to the
vector multiplets that are either abelian or
in the Cartan of the non-abelian gauge groups.
There are $h^{2,1}(\Xh)+1$ massless hypermultiplets in the
five-dimensional theory. In the decompactification limit,
they are hypermultiplets uncharged under the Cartan/abelian
vector multiplets.
There are $(\sum_r 1 + \sum_\rho 2 g_\rho)$ massive hypermultiplets
and $(\sum_\rho 2)$ massive vector multiplets, which in the decompactification limit,
are hyper/vector multiplets charged under the abelian or non-abelian Cartan
vector multiplets.

Since we have accounted for all the vector and hypermultiplets
of the six-dimensional theory from the geometric data of $\Xh$,
the gravitational anomaly constraint
\be
H-V+29T=273 \,,
\ee
can be written in terms of this data.
The number of six-dimensional vector multiplets and
hypermultiplets are given by
\begin{align}
V&=(h^{1,1} (\Xh)-2-T)+\sum_\rho 2 \\
H&=(h^{2,1} (\Xh)+1)+\sum_r 1 + \sum_\rho 2g_\rho \,.
\end{align}
Thus, the gravitational anomaly constraint
can be re-written as
\be
270-30T+(h^{1,1}(\Xh) -h^{2,1}(\Xh))= \sum_r 1 + \sum_\rho (2g_\rho -2) \,.
\ee
Using the fact that $K \cdot K = 9-T$ for the canonical divisor $K$
of $\BB$, and that $\chi_\Xh = 2(h^{1,1}(\Xh) -h^{2,1}(\Xh))$ for the
Euler characteristic $\chi_\Xh$ of $\Xh$,
we find
\be
30K \cdot K + {1 \ov 2} \chi_\Xh =  \sum_r 1 + \sum_\rho (2g_\rho -2) \,.
\ee

\subsubsection{The Non-Abelian Sector} \label{sss:nonab}

In this section, we continue the analysis of the M-theory/F-theory
duality. We first classify the vector fields of the five-dimensional
theory in a useful way. Then we recover the low-energy data of the
non-abelian sector from the geometric data of $\Xh$.
The results turn out to be consistent with that of section \ref{ss:revKMT}.
Most of the discussion in this section
can be found in \cite{AKM,GM2000,GM2011,BIKMSV,KV}
but we have rephrased them in a way more convenient
for our purposes.

Let us first classify the vector fields of the five-dimensional theory
in a useful way. Recall there is a one-to-one correspondence
between the massless five-dimensional vector fields
and four-cycles of $\Xh$.
There are four types of four-cycles in $\Xh$.
\begin{enumerate}
\item \textbf{Type \^Z : }The zero section; $\hat{Z}$ $\leftrightarrow \hat{\zeta}$.
\item \textbf{Type B : }Four-cycles obtained by fibering the elliptic fiber
$f$ over two-cycles $H_0, \cdots, H_{T}$
in the base $\BB$; $B_0, \cdots, B_{T}$ $\leftrightarrow \beta_0, \cdots, \beta_T$.
\item \textbf{Type C : }Monodromy invariant four-cycles that are locally type F rational curves
fibered over a curve in the base $\BB$; $T_1, \cdots, T_r$ $\leftrightarrow \tau_1, \cdots, \tau_R$.
\item \textbf{Type \^S : }Non-zero sections of the fibration;
$\Sh_1, \cdots, \Sh_\n$ $\leftrightarrow \hat{\sigma}_1, \cdots, \hat{\sigma}_\n$.
\end{enumerate}
The lowercase greek letters denote the Poincar\'e dual two-forms
in the resolved manifold.
The type \^S four-cycles are generators of the non-torsion part of the
Mordell-Weil group of the elliptic fibration. The number $\n$ is the Mordell-Weil
rank of the elliptic fibration \cite{MV2, Wazir,HulekKloosterman}.

We note that the intersection of type B cycles satisfy
\be
B_\alpha \cdot B_\beta = (H_\alpha \cdot H_\beta )_\BB f \equiv \Om_{\alpha \beta} f \,,
\label{BB}
\ee
where the subscript $\BB$ means that we are taking
the intersection of curves on the base and
$f$ is the fiber class of the elliptic fibration.
We emphasize once more that
$H_\alpha$ are the basis elements of $H_2 (\BB)$.
From this relation we also see that the triple intersection products
among the type B cycles are zero, as the 
type B cycles do not intersect a generic fiber.
The $\Om$ is a symmetric invertible $SO(1,T)$ bilinear form,
or an $SO(1,T)$ ``metric."
We denote
\be
\Om^{\alpha \beta} \equiv (\Om^{-1})_{\alpha \beta}
\ee
and raise and lower $SO(1,T)$ indices by $\Om.$

We postpone the discussion of the four cycles of type \^S
to section \ref{sss:ab} and focus on the first three types of cycles.
We make the following
\\

\noindent
\textbf{(Claim 1)}
\ben
\item The vector field $Z$ obtained by the three-form KK-reduced
along $\zeta=\hat{\zeta}-{1 \ov 2} (\hat{Z} \cdot \hat{Z} \cdot B^\alpha ) \beta_\alpha$
can be identified with the vector field coming from KK-reducing the six-dimensional metric
along $S^1$ in the decompactification limit.  It is inside the five dimensional gravity multiplet.
\item The vector fields $B_0,\cdots,B_T$ obtained by
the three form KK-reduced along $\beta_0, \cdots, \beta_T$
can be identified with the vector fields obtained by KK-reducing the $(T+1)$
six-dimensional tensor fields along $S^1$ in the decompactification limit.
\item The vector fields $\sA_{1}, \cdots, \sA_{R}$ obtained by the
three-form KK-reduced along $\tau_i$ can be identified with
the vector fields obtained by KK-reducing the
six-dimensional non-abelian vector fields
in the coroot basis of the Cartan of each gauge group
along $S^1$ in the decompactification limit.
\een
For convenience, we abuse the term ``duality" throughout
the course of the paper in the following way;
we say that a vector field is ``dual to" a four-cycle $S$ when it is
obtained by KK-reducing the eleven-dimensional
three-form on a two-form that is Poincar\'e dual to $S$.

We denote the Poincar\'e dual four-cycle of $\zeta$ as
\be
Z=\hat{Z}-{1 \ov 2} (\hat{Z} \cdot \hat{Z} \cdot B^\alpha ) B_\alpha
\ee
$Z$ has been defined so that
\be
B_\alpha \cdot Z \cdot Z =0
\ee
for all $\alpha$, as can be checked explicitly.
Also, as the $B_\alpha$ do not intersect the fiber, $Z \cdot f = \hat{Z} \cdot f =1$.
We denote this four-cycle a \textbf{type Z} four-cycle.

Let us verify the third entry of (Claim 1) first.
We can always organize the basis of
type C cycles in a convenient way.
For each non-abelian gauge group $\GG_\kappa$,
there exists a curve $b_\kappa$ in the base over which
the fiber takes the Kodaira fiber-type corresponding to $\GG_\kappa$.
We use $b_\kappa$ to denote both the actual curve and its class
in the base.
The fiber at $b_\kappa$ consists of groups of type F rational curves.
There is a canonical way of choosing linearly independent
monodromy invariant groups of these rational curves,
which we discuss in length in appendix \ref{ap:lieint}.
If we denote this group as $\gamma_{I,\kappa}$ for each $\kappa$,
the four cycles obtained by fibering $\gamma_{I,\kappa}$ over $b_\kappa$
are dual to the vector field corresponding to $\sT_{I,\kappa}$,
the $I$'th element of the coroot basis of the Cartan of $\GG_\kappa$.
This is because, as we have checked in appendix \ref{sap:int},
the intersection numbers reproduce the charges of the charged adjoint multiplets
under $\sT_{I,\kappa}$ correctly.

To be more precise, let us denote $T_{I,\kappa}$ to be
the four cycle obtained by fibering $\gamma_{I,\kappa}$
over the curve $b_\kappa$.
As checked case by case for each Lie group in \ref{sap:int},
for each $\kappa$ we can find type F rational curves that correspond
to the simple roots $\alpha_{I,\kappa}$ of the Lie group $\GG_\kappa$.
Let us denote those curves $\chi_{I,\kappa}$. Then,
\be
T_{I,\kappa} \cdot \chi_{J,\lambda} =
-\delta_{\kappa\lambda}
{2 \vev{\alpha_{I,\kappa}, \alpha_{J,\kappa} } \ov \vev{\alpha_{I,\kappa}, \alpha_{I,\kappa}} }
= -\delta_{\kappa\lambda} C_{IJ,\kappa} \,,
\ee
where $C_\kappa$ is the Cartan matrix of $\GG_\kappa$.
All type F rational curves $\chi_\rho$ can be written as
linear combinations of $\chi_{I,\kappa}$.
The intersection numbers between these curves and $T_{I,\kappa}$
precisely reproduce the charges all the negative roots of each gauge group.
By wrapping branes and anti-branes along these type F curves,
one recovers all the charged adjoint vector fields of the theory.
In the F-theory limit, these charged vector fields become massless,
and along with the vector fields dual to $T_{I,\kappa}$ form the
vector multiplet of gauge group $\GG_\kappa$.

Therefore we can group
the type C cycles according to their gauge groups
{\it i.e.,}
\be
T_{I,\kappa}~:~(T_{1,1}, \cdots, T_{R_1,1}),\cdots,(T_{1,N},\cdots,T_{R_{N},{N}})\,.
\ee
These are dual to non-abelian gauge field components
\be
\sA_{I,\kappa}~:~(\sA_{1,1}, \cdots, \sA_{R_1,1}),\cdots,(\sA_{1,N},\cdots,\sA_{R_{N},{N}})
\ee
of the coroot basis elements of the Cartans of the non-abelian gauge groups;
\be
\sT_{I,\kappa}~:~(\sT_{1,1}, \cdots, \sT_{R_1,1}),\cdots,(\sT_{1,N},\cdots,\sT_{R_{N},{N}})\,.
\ee

Meanwhile, hypermultiplets obtained by wrapping M2 branes around
clusters of type I curves in the resolved manifold
also form representations under the non-abelian gauge groups.
The representations can be determined by
computing the intersection number of all the type I curves
in each ``cluster" with each $T_{I,\kappa}$.
Note that the intersection number between
$T_{I,\kappa}$ and any rational curve is integral.
This is consistent with the fact that
the charge of any weight vector is integral under elements of the
coroot basis.

There is one question we raise before we go further.
The type B cycles and type Z cycles do not
intersect any of the shrinking curves, {\it i.e.,}
\be
B_\alpha \cdot c_r = 
B_\alpha \cdot \chi_\rho = 
Z \cdot c_r = 
Z \cdot \chi_\rho =0 \,.
\ee
A given shrinking rational curve sits at a point above the base,
and hence $B_\alpha$, which is a fibration over a curve in the base,
can always be smoothly deformed to avoid intersecting it.
$Z$ does not intersect any shrinking curves since
the zero section does not touch any singularities in the fibration limit.
Therefore, a vector field obtained by reducing the 11D three-form
over some type C cycle $T$, and a vector field dual to
$T+xZ+t^\alpha B_\alpha$
should reproduce the same charges.
We claim that $x$ and $t^\alpha$ can be fixed to zero
by comparing the coefficient for the Chern-Simons
five-form for the vector fields.

We now justify the two claims about vector fields $Z$ and $B^\alpha$
by examining the Chern-Simons term in the
five-dimensional theory in a generic point in the Coulomb branch.
It is given by \cite{FerraraKhuriMinasian}
\be
(\SS_x \cdot \SS_y \cdot \SS_z) \int A^x \wedge F^y \wedge F^z \,,
\ee
where $\SS_x$ are the dual four-cycles of the two-forms
each gauge field is KK-reduced upon.
The coefficient is the triple intersection of the four-cycles
involved.

The intersection numbers are given by
\begin{align}
\begin{split}
&{1 \ov 4}(9-T)(ZZZ)
+{3\ov 2} \delta_{\kappa\lambda} \CC_{IJ,\kappa} (K \cdot b)_\BB (Z T_{I,\kappa} T_{J,\lambda}) \\
&+ 3 \Om_{\alpha\beta}(ZB_\alpha B_\beta)
-3 \delta_{\kappa\lambda} \CC_{IJ,\kappa} b_{\alpha,\kappa} (B_\alpha T_{I,\kappa} T_{J,\lambda}) \\
&+\text{(triple intersections among $T$'s)}
\end{split}
\label{intnonab}
\end{align}
in standard polynomial notation\footnote{We have not computed the triple
intersections among the type T cycles, as we do not need them for the purposes
of this paper. We note that these terms have been computed and matched
with the F-theory side in \cite{GrimmBonetti}.};
the coefficient of the term $(\SS_x \SS_y \SS_z)$
is the intersection number $(\SS_x \cdot \SS_y \cdot \SS_z)$ with multiplicities, {\it i.e.,}
the polynomial is defined as
\be
\sum_{x,y,z} (\SS_x \cdot \SS_y \cdot \SS_z) (\SS_x \SS_y \SS_z)\,.
\ee
$b^{\alpha}_\kappa$ are the $SO(1,T)$ coordinates of $b_\kappa$, {\it i.e.,}
\be
b_\kappa = b^\alpha_\kappa H_\alpha \,.
\ee
$\CC_\kappa$ is the normalized coroot inner-product matrix
for gauge group $\GG_\kappa$
defined in section \ref{ss:factcond} and discussed extensively
in appendix \ref{ap:lieint}.
$K$ is the canonical divisor of the base.

Let us explain this result first.
It is more convenient to obtain the intersection numbers using $\Zh$,
so we use the $\Zh$ rather than $Z$.
Intersection numbers involving $Z$ can be obtained straightforwardly
from those involving $\Zh$.

We first note that
\be
\Zh \cdot X \cdot Y = \left( X|_\BB \cdot Y|_\BB \right)_\BB
\ee
since $\Zh$ is the normal bundle of the base $\BB$.
$X|_\BB$ is the two-cycle on $\BB$---or more precisely,
the zero section $\Zh$---obtained by restricting
the four-cycle $X$ to $\Zh$.
The manifold $\Xh$ is Calabi-Yau and hence
by the adjunction formula
\be
\Zh |_\BB = K \,.
\ee
Also,
\be
B_\alpha |_\BB = H_\alpha \,.
\ee
Hence
\begin{align}
\begin{split}
\Zh \cdot \Zh \cdot \Zh = (K \cdot K)_\BB = (9-T)&, \quad
\Zh \cdot \Zh \cdot B_\alpha = (K\cdot H_\alpha)_\BB = K_\alpha, \\
\Zh \cdot B_\alpha \cdot B_\beta &= (H_\alpha \cdot H_\beta)_\BB = \Om_{\alpha\beta} \,.
\end{split}
\end{align}

By construction, $\Zh$ and $T_{I,\kappa}$ are disjoint;
the zero section does not touch any of the singularities.
Hence for any four cycle $X$,
\be
\Zh \cdot T_{I,\kappa} \cdot X =0 \,.
\ee
Also, by \eq{BB}
\be
B_\alpha \cdot B_\beta \cdot T_{I,\kappa}
= \Om_{\alpha\beta} f \cdot T_{I,\kappa} \,.
\ee
Since $T_{I,\kappa}$
are rational curves fibered along a curve in the base,
it does not intersect with a generic fiber.
Hence the intersection number is $0$.

It is shown in appendix \ref{ap:lieint} that 
\be
B_\alpha \cdot T_{I,\kappa} \cdot T_{J,\lambda}
= -b_{\alpha,\kappa} (\gamma_{I,\kappa} \cdot T_{J,\lambda})
=- \delta_{\kappa\lambda} \CC_{IJ} b_{\alpha,\kappa} \,.
\label{BTT}
\ee
$\kappa$ and $\lambda$ must be the same in order
to get a non-zero result since $T_{I,\kappa}$ does not
intersect type F rational curves fibered over
a different locus.

It is convenient to express this relation using
the projection $\pi$ to the base manifold.
More precisely, we define $\pi(C)$ of some two-cycle $C$
in $\Xh$ to be the projection of
$C$ to the $H_2(\BB)\cong H^{1,1}(\BB)$ lattice of the
base manifold.
As pointed out in the introduction, the projection of
$C$ to the base can in general be a linear combination
of two, one and zero-cycles in the base.
We treat one-cycles and zero-cycles to be
null vectors in $H_2(\BB)$.
Then $\pi$ is defined so that for any two-cycle $C$
in $\Xh$,
\be
\pi(C) = (C \cdot B^\alpha) H_\alpha
\quad \Leftrightarrow \quad
B_\alpha \cdot C = (H_\alpha \cdot \pi(C))_\BB = 
\pi(C)_\alpha \,.
\ee
Therefore \eq{BTT} can be rewritten as
%$T_{I,\kappa}$ and $T_{J,\lambda}$ meet at exactly
%`$- \delta_{\kappa\lambda} \CC_{IJ}$ points'
%over the curve of class $b_\kappa$. Since $B_\alpha$
%is given by the elliptic fiber fibered over $H_\alpha$
\be
\pi( T_{I,\kappa} \cdot T_{J,\lambda})
=- \delta_{\kappa\lambda} \CC_{IJ} b_{\alpha,\kappa} H^\alpha
=- \delta_{\kappa\lambda} \CC_{IJ} b_{\kappa} \,.
\ee

Now let us investigate the six-dimensional F-theory
background compactified on the
singular manifold $X$ and then further compactified on $S^1$.
Let us denote the vector fields obtained by KK-reduction on $S^1$
in the following way:
\ben
\item $Z'$ is the vector field obtained by KK-reducing the six-dimensional metric. It is inside the gravity multiplet.
\item $B'$ are the vector fields obtained by KK-reducing the $(T+1)$ tensors.
\item $\sA'$ are the vector fields obtained by KK-reducing the non-abelian vector fields in the
coroot basis of the Cartan of the gauge group.
\een
Let us denote the anomaly coefficients for the
non-abelian gauge fields as $b_\kappa'$.
In section 3 of \cite{FerraraMinasianSagnotti}, the coefficients
of the Chern-Simons term of the KK-reduced
five-dimensional theory on a generic point in the Coulomb branch
is worked out explicitly.
The intersection polynomial is given by
\begin{align}
\begin{split}
&\Om_{\alpha\beta}(Z'B'_\alpha B'_\beta)
- 2\delta_{\kappa\lambda} \CC_{IJ,\kappa} b'_{\alpha,\kappa} (B'_\alpha \sA'_{I,\kappa} \sA'_{J,\lambda}) \\
&+\text{(triple intersections among $\sA$'s)}
\end{split}
\label{intsix}
\end{align}
up to an overall constant---that we denote $K_{int}$---in the
``decompactification limit," {\it i.e.,} when the vacuum expectation value
of the scalars in the vector multiplets and the inverse radius
of the $S^1$ go to zero \cite{FerraraMinasianSagnotti}.
Note that we have used the non-trivial fact that $\lambda_\kappa$ is chosen
so that the Cartan generators $\{ \sT_{I,\kappa} \}$ of $\GG_\kappa$ in the
coroot basis satisfy
\be
{1 \ov \lambda_\kappa}\tr \sT_{I,\kappa} \sT_{J,\kappa} = \CC_{IJ,\kappa} \,.
\ee

This intersection form agrees with \eq{intnonab} up to terms that do not
involve $B$ when we identify $b_\kappa = b'_\kappa$---which is
indeed true for non-abelian gauge fields \cite{KMT2, KMT1}---and
take $Z$ and $B_\alpha$ to be proportional to $Z'$ and $B_\alpha'$.
The terms that involve $B$ cannot receive corrections for the following
reason. The corrections to these Chern-Simons terms come from
one-loop integrals of five-dimensional fermions \cite{WittenPT}.
The only way that terms involving $B$ could receive corrections
on the F-theory side is if some six-dimensional fermion in a short multiplet
couples to the tensor field $B$ in a way that reduces to
\be
\bar{\psi} B_\mu \Gamma^\mu \psi
\ee
in five dimensions. There are no such couplings so these terms are not
modified \cite{GrimmBonetti}.
Meanwhile, the vector field $Z_\mu \sim g_{\mu 5}$ can couple in this
manner to charged fermions in short multiplets.
One-loop contributions of these fermions
generate the first two terms of \eq{intnonab}
\cite{GrimmBonetti}.

We note that we have a well defined normalization
prescription for $Z$ and $B$ given in the following way.
There is an unambiguous prescription for the normalization
of non-abelian gauge fields on both sides;
they were normalized to reproduce the charges of the coroot lattice.
This implies that $\sA$ and $\sA'$ are indeed identical.
Then we can fix the proportionality constant of the $B$
with respect to $B'$ by using the fact that
\be
b_\kappa = b'_\kappa \,.
\ee
This in turn fixes the proportionality constant of $Z$
with respect to $Z'$.
Fixing the normalization of $B$ and $Z$ is important in
determining what the abelian anomaly coefficients are.

We have verified (Claim 1) by comparing the Chern-Simons
five form on the M-theory and F-theory side.
Note that if we add type B or type Z cycles to type C cycles,
the intersection polynomial becomes modified.
In particular, terms of form $(ZBT)$ would appear,
which do not and cannot appear on the F-theory side
in the decompactification limit.

\subsubsection{The Abelian Sector} \label{sss:ab}

In this section we find the four-cycles dual to
KK-reduced abelian gauge fields and identify the
abelian anomaly coefficients.
We make the following

\medskip

\noindent
\textbf{(Claim 2)} The vector fields $A_1, \cdots, A_\n$ dual to
type S four-cycles $S_i$---which we shortly define---can be identified with
the vector fields obtained by KK-reducing the six-dimensional abelian
vector fields along $S^1$ in the decompactification limit.

\medskip

We construct the \textbf{type S} four-cycles in the following way.
For each type \^S four-cycle $\Sh_i$, define the corresponding type S
four-cycle $S_i$ as
\be
S_i = \Sh_i - (\Sh_i \cdot f) \Zh -((\Sh_i - (\Sh_i \cdot f) \Zh) \cdot \Zh \cdot B^\alpha) B_\alpha
+ \sum_{I,J,\kappa} (\Sh_i \cdot \chi_{I,\kappa}) (C_\kappa^{-1})_{IJ} T_{J,\kappa}
\label{Shioda}
\ee
where $\kappa$ labels the non-abelian gauge groups of the
six-dimensional theory and $I$ labels
their simple roots. $(C_\kappa^{-1})_{IJ} $ is the
$(I,J)$ component of the inverse of the Cartan matrix $C_{\kappa}$.
Recall that $\chi_{I,\kappa}$ is the type F cycle corresponding to the
simple roots $\alpha_{I,\kappa}$ of $\GG_\kappa$.
Type S cycles are defined so that
\ben
\item $S_i \cdot f =0$.
\item $S_i \cdot \Zh \cdot B_\alpha =0$.
\item $S_i \cdot \chi_\rho =0$.
\een
The first and second identities can be checked easily
by using intersection identities given in the previous section.
We note that the first condition implies that
\be
S_i \cdot B_\alpha \cdot B_\beta
= \Om_{\alpha\beta} S_i \cdot f 
= 0\,.
\ee
We also note that the second condition implies that
\be
S_i \cdot \Zh \cdot X
= (S_i|_\BB \cdot X|_\BB )_\BB
= (X|_\BB)^\alpha S_i \cdot \Zh \cdot B_\alpha =0
\ee
for any four-cycle X.

Since all $\chi_\rho$ are
homologically equivalent to a sum of $\chi_{I,\kappa}$,
the third identity needs to be checked only for all $\chi_{I,\kappa}$.
This can be done:
\begin{align}
\begin{split}
S_i \cdot \chi_{I,\kappa}
&= \Sh_i \cdot \chi_{I,\kappa}
+ \sum_{J,K,\lambda} (\Sh_i \cdot \chi_{J,\lambda}) (C_\lambda^{-1})_{JK} (T_{K,\lambda} \cdot \chi_{I,\kappa}) \\
&=\Sh_i \cdot \chi_{I,\kappa}
+ \sum_{J,K,\lambda} (\Sh_i \cdot \chi_{J,\lambda}) (C_\lambda^{-1})_{JK} (-\delta_{\lambda\kappa} C_{KI,\kappa}) \\
&=\Sh_i \cdot \chi_{I,\kappa}
-  \Sh_i \cdot \chi_{I,\kappa} =0 \,.
\end{split}
\end{align}

Meanwhile,
\begin{align}
\begin{split}
S_i \cdot T_{I,\kappa} \cdot B_\alpha 
= b_{\alpha,\kappa} (S_i \cdot \gamma_{I,\kappa}) \,.
\end{split}
\end{align}
Recall that $\gamma_{I,\kappa}$ is the monodromy invariant fiber of
$T_{I,\kappa}$ over $b_{\kappa}$.
As can be seen in appendix \ref{ap:lieint},
the $\gamma_{I,\kappa}$ are linear combinations of $\chi_{J,\kappa}$.
Therefore, it follows that
\be
S_i \cdot T_{I,\kappa} \cdot B_\alpha =0
\ee
for any $T_{I,\kappa}$ and $B_\alpha$.

Equation \eq{Shioda} is the threefold analog of the map Shioda
used to map rational sections of an elliptically fibered surface
to points in the N\'eron-Severi lattice of that surface
\cite{Shioda}.\footnote{The image of the
rational sections of an elliptically fibered K3 manifold
$\mathcal{M}$ under the Shioda
map---which are two-cycles---are also dual
to the abelian vector fields of the eight-dimensional
supergravity theory obtained by compacifying F-theory
on $\mathcal{M}$ \cite{VafaF,AspinwallK3,FYY,Guralnik}.}
Once the rational sections were mapped to a lattice,
a number-valued inner-product on the sections
could be defined.
In our case there is an $H_2(\BB)$ vector-valued
inner-product on type S cycles.
It is $-\pi(S_i \cdot S_j)$. We claim that these are
the anomaly coefficients of the abelian gauge groups.

Now we verify that vector fields dual to type S cycles
can indeed be identified with the
KK-reduced six dimensional abelian vector fields
in the decompactification limit.
We verify that none of the vector multiplets are charged
under $A_i$ and that the coefficients of the Chern-Simons
five-forms have the proper form.
The first point is easily checked since
$S_i \cdot \chi_\rho =0$ implies that none of the
charged vector multiplets are charged under $A_i$.
$S_i$, however, can intersect with curves of type I,
{\it i.e.,} hypermultiplets can be charged under
abelian gauge fields.

The triple intersection polynomial, when incorporating
$S_i$ becomes
\begin{align}
\begin{split}
&{1 \ov 4}(9-T)(ZZZ)
+{3\ov 2} \delta_{\kappa\lambda} \CC_{IJ,\kappa} (K \cdot b)_\BB (Z T_{I,\kappa} T_{J,\lambda})
-{3\ov 2} (K \cdot \pi(S_i \cdot S_j))_\BB (Z S_i S_j)\\
&+3 \Om_{\alpha\beta}(ZB_\alpha B_\beta)
- 3 \delta_{\kappa\lambda} \CC_{IJ,\kappa} b_{\alpha,\kappa} (B_\alpha T_{I,\kappa} T_{J,\lambda})
+ 3 \pi(S_i \cdot S_j)_\alpha (B_\alpha S_i S_j)\\
&+\text{(triple intersections among $T$, $S$)}
\end{split}
\label{intab}
\end{align}
We have explained the absence of the terms $(SBB)$,
$(SBZ)$, $(STZ)$ and $(STB)$.
Coefficients of the $(SSB)$ terms follow from the definition of
the projection $\pi$.
%\be
%B_\alpha \cdot S_i \cdot S_j
%= \pi(S_i \cdot S_j)_\alpha \,.
%\ee

Adding the contributions of $A'$---the vector fields obtained by
KK-reducing the six-dimensional abelian vector fields---to equation
\eq{intsix}, the tree-level intersection polynomial on the F-theory
side is given by
\begin{align}
\begin{split}
&\Om_{\alpha\beta}(Z'B'_\alpha B'_\beta)
-2\delta_{\kappa\lambda} \CC_{IJ,\kappa} b'_{\alpha,\kappa}
(B'_\alpha \sA'_{I,\kappa} \sA'_{J,\lambda})
- 2b_{\alpha, ij} (B'_\alpha A'_{I,\kappa} A'_{J,\lambda}) \\
&+\text{(triple intersections among $\sA$'s and $A$'s)}
\end{split}
\label{intsixab}
\end{align}
up to the same overall constant $K_{int}$ defined below \eq{intsix}.
Recall that $b_{\alpha,ij}$ are the $SO(1,T)$ vector
components of the abelian anomaly coefficients.
We see that the intersection polynomial \eq{intab} matches
with \eq{intsixab} up to terms not involving $B$,
when we normalize $Z$ and $B$ with respect to $Z'$ and $B'$
according to the prescription given at the end of the previous section.
The matching of the intersection polynomial concludes
the justification of (Claim 2).

Furthermore, if we normalize the gauge fields $A'_i$
so that the charge of the
hypermultiplet coming from branes wrapping $c_r$
is $c_r \cdot S_i$, we can equate $A'_i$ and $A_i$.
Then due to the normalization
prescription of $B_\alpha$
we have given in the previous section,
we can unambiguously equate
\be
\label{mainresult}
b_{ij} = -\pi(S_i \cdot S_j)\,.
\ee
This is the main result of this section.

\subsection{Summary} \label{ss:summ}

Let us summarize our findings.
F-theory compactified on $X\times S^1$---where $X$
is an elliptically fibered Calabi-Yau threefold with a section---is
dual to M-theory compactified on $X$.
We have identified the massless field content of the
six-dimensional theory from the M-theory dual.
In the process, we have proven equation \eq{three}.

The vector fields of the six-dimensional
theory KK-reduce along $S^1$ to
vector fields in five dimensions.
In the M-theory dual, the KK-reduced vector fields
have the following origins:
\begin{enumerate}
\item \textit{Abelian Vector Fields : }KK-reduction of the 11D
three-form on two-forms dual to four-cycles of type S.
\item \textit{Non-abelian Vector Fields in the Cartan of
the Gauge Group: }
KK-reduction of the 11D three-form on
two-forms dual to four-cycles of type C.
\item \textit{Non-abelian Vector Fields Not in the
Cartan of the Gauge Group: }
M2 branes and anti-branes wrapping curves of type F.
\end{enumerate}
The definitions of the various types of cycles are
given in section \ref{ss:Mthy}.
We note once again that we abuse the term ``duality" in the
following way;
we say that a vector field is ``dual to" a four-cycle $\SS$
when it is obtained by KK-reducing the eleven-dimensional
three-form on a two-form Poincar\'e dual to $\SS$.
The abelian and non-abelian Cartan vector multiplets
are dual to four-cycles that do not intersect the fiber.

We elaborate on the construction of cycles of type S.
Type S cycles $S_i$ are constructed from four-cycles that are the
generators of the rational sections through the Shioda map \eq{Shioda}.
The anomaly coefficient of the abelian vector fields can be identified
with the opposite vector of the projection of the
the intersection of two type S four-cycles to the
$H^{1,1}$ lattice of the base:
\be
b_{ij} = - \pi (S_i \cdot S_j) \,.
\ee

All the fields charged under abelian or non-abelian Cartan vector fields
come from M2 branes wrapping shrinking rational curves.
Rational curves of type I---or isolated rational curves---contribute
one hypermultiplet each to the massless spectrum
in the decompactification limit:
a brane and an anti-brane wrapping a given curve
contribute a half-hypermultiplet each, which together
form one hypermultiplet.
Rational curves of type F---or fibered rational curves---contribute
$2g$ hypermultiplets where $g$ is the genus of the curve
over which the rational curve is fibered.
As mentioned above, a type F rational curve also contributes
two vector multiplets to the massless spectrum of the six-dimensional theory,
each obtained by either wrapping a brane or an anti-brane.

A charged hypermultiplet consists of two half-hypermultiplets
each coming from wrapping an M2 brane or anti-brane on a curve.
There is an overall sign ambiguity in defining charges of hypermultiplets.
We use the convention that a hypermultiplet coming from wrapping
branes and anti-branes on a rational curve $C$ has charge $C \cdot \SS$
under the vector field dual to a four-cycle $\SS$.
Meanwhile, each vector field coming from wrapping M2 branes(anti-branes)
on the type F curve $\chi_\rho$ has charge
$\chi_\rho \cdot \SS$($-\chi_\rho \cdot \SS$)
under the vector multiplet dual to a four-cycle $\SS$, respectively.

\section{Anomaly Equations and Intersection Theory} \label{s:intersection}

Due to the identifications made in the previous section,
the mixed/gauge anomaly equations can be reformulated into
equalities between intersection numbers obtained in
the resolved Calabi-Yau threefold.
Remarkably, they can be summarized in two equalities.
They are given by the following:

\begin{align}
\begin{split}
&\pi(\SS_1 \cdot \SS_2)\cdot \pi(\SS_3 \cdot \SS_4)+
\pi(\SS_1 \cdot \SS_3)\cdot \pi(\SS_2 \cdot \SS_4)+
\pi(\SS_1 \cdot \SS_4)\cdot \pi(\SS_2 \cdot \SS_3) \\
&=\sum_r (c_r \cdot \SS_1) (c_r \cdot \SS_2) (c_r \cdot \SS_3) (c_r \cdot \SS_4)
+ \sum_\rho (2g_\rho-2)(\chi_\rho \cdot \SS_1) (\chi_\rho \cdot \SS_2) (\chi_\rho \cdot \SS_3) (\chi_\rho \cdot \SS_4)
\end{split}
\label{1}
\end{align}
and
\begin{align}
\begin{split}
6K \cdot \pi(\SS_1 \cdot \SS_2) 
=\sum_r (c_r \cdot \SS_1) (c_r \cdot \SS_2) 
+ \sum_\rho (2g_\rho-2)(\chi_\rho \cdot \SS_1) (\chi_\rho \cdot \SS_2) 
\end{split}
\label{2}
\end{align}
when
\be
f \cdot \SS_n =0\,.
\ee
As in the previous section, $c_r$ denote the rational curves of type I,
while $\chi_\rho$ denote the rational curves of type F.
Recall that by definition $c_r$ and $\chi_\rho$ are curves that
shrink to zero area in the fibration limit.
$g_\rho$ denotes the genus of the curve over which rational curve
$\chi_\rho$ is fibered. We have used $K$ to denote the canonical
class divisor of the base.

$\pi$ is the projection to the base manifold.
More precisely, $\pi(C)$ of some two-cycle $C$
in $\Xh$ is the projection of $C$ to the $H_2(\BB)$
lattice of the base manifold.
The intersection between projected curves are taken in the base,
while all the other intersections are taken inside the full manifold.
%We note that $\pi(\SS_n \cdot \SS_m)\cdot \pi(\SS_k \cdot \SS_l)$
%is non-zero only when both $\pi(\SS_n \cdot \SS_m)$ and
%$\pi(\SS_k \cdot \SS_l)$  are two-cycles on the base, as
%$\pi(\SS_n \cdot \SS_m)$ is at most of dimension two.
Recall that for any two-cycle $C$
in $\Xh$
\be
\pi(C) = (C \cdot B^\alpha) H_\alpha
\quad \Leftrightarrow \quad
B_\alpha \cdot C = (H_\alpha \cdot \pi(C))_\BB = 
\pi(C)_\alpha \,,
\ee
for the basis elements $H_\alpha$ of $H_2(\BB)$.

As seen in section \ref{ss:Mthy}, any four-cycle
that does not intersect the
fiber is a linear combination of four-cycles of type B, S, or C.
One can easily check that to prove equations \eq{1} and \eq{2} for
any four-cycle with zero intersection with the fiber,
it is enough to prove them in the case when all $\SS_n$
are among the basis elements $\{ B_\alpha, T_{I,\kappa}, S_i \}$.
We can carry out this procedure in the following steps.
\ben
\item We first show that these equations trivially hold when one
of $\SS_n$ is of type B.
\item We then show that these equations hold when all four four-cycles
$\SS_n$ are of type S or of type C.
\item Finally we show the validity of the equations
when there are both four-cycles of type S and C
among $\SS_n$ thereby concluding the proof of these equations.
\een

The details of these steps are unilluminating,
but the basic idea is simple.
For the rest of the section, we carry out step 1 explicitly
and sketch the idea behind showing steps 2 and 3.
We have carried out steps 2 and 3 explicitly in appendix \ref{ap:2and3}.

Let us prove equations \eq{1} and \eq{2} in the case
that one of the four cycles is of type B.
Without loss of generality, let $\SS_1 = B_\alpha$.
All shrinking two-cycles do not intersect $\SS_1$.
Therefore the right-hand sides of both equations are 0.
Meanwhile, for any $S$ such that $S \cdot f =0$
\be
B_\alpha \cdot B_\beta \cdot S = \Om_{\alpha\beta} f \cdot S = 0
\ee
for all $B_\beta$ and therefore
\be
\pi(B_\alpha \cdot S) = (B_\beta \cdot B_\alpha  \cdot S) H^{\beta} =0 \,.
\ee
Therefore $\pi (\SS_n \cdot \SS_1)=0$ for $n=2,3,4$
and hence the left-hand sides of the two equations are also zero.

When all $\SS_n$ are either of type C or S,
the equations \eq{1} and \eq{2} become more interesting.
In this case, the gauge anomaly equations \eq{gauge}
lead to \eq{1} and the mixed anomaly equations \eq{mixed}
lead to \eq{2}.
As can be seen in section \ref{ss:Mthy}, each
gauge field $A_x$ in the Cartan subalgebra of the full gauge group
is dual to a four-cycle $C_x \in \{ T_{I,\kappa}, S_i \}$
in the resolved Calabi-Yau manifold $\Xh$.
If we restrict our attention to only these gauge fields,
the anomaly polynomial takes the structure of an abelian theory.
In particular, the anomaly coefficients of $F_x F_y$ are given by
$-\pi(C_x \cdot C_y)$---they are bilinear forms
in the $x$ index and
are vectors in the $H_2$ lattice of the base.
Therefore by plugging in elements of the Cartan
to the gauge/mixed anomaly equations,
the inner-products between anomaly coefficients on
left-hand sides reproduce the intersection
numbers between between various $\pi(\SS_i \cdot \SS_j)$
of \eq{1} and \eq{2}.

The right-hand sides of the gauge/mixed anomaly equations
\eq{gauge}/\eq{mixed},
are given by the sum of products of the charges of
``charged multiplets" under vector fields dual to $C_x \in \{ T_{I,\kappa}, S_i \}$.
The charged multiplets come from quantizing zero-modes
of the M2 branes and anti-branes wrapping type I or type F curves.
A type I curve $c_r$ contributes one hypermultiplet with
charge $c_r \cdot C_x$, while a type F curve $\chi_\rho$
contributes $2g_\rho$ hypermultiplets of charge $\chi_\rho \cdot C_x$
and two vector multiplets each with charge $\pm \chi_\rho \cdot C_x$
under the vector field dual to $C_x$ \cite{WittenPT,KMP}.
Therefore the right-hand sides
of equations \eq{1}/\eq{2} are reproduced by
plugging in elements of the Cartan to the right-hand sides
of the gauge/mixed anomaly equations.
This concludes the proof of the two equations.

\section{Summary and Discussion} \label{s:discussion}

Using anomaly equations we have presented a physics
proof of the equations \eq{one}, \eq{two} and \eq{three}
given in the introduction.
The geometric implications of the mixed and
gauged anomaly equations
have been studied in \cite{GM2000,GM2011} for non-abelian
gauge groups, but have not been put into the form
we have presented in \eq{one} and \eq{two}.
The implications of the third equation---coming from
the gravitational anomaly constraint---has also been studied previously
\cite{AKM,GM2000,GM2011},
although not quite in the language that we have used.
An interesting fact is that the equation \eq{three}
can be translated into a threefold analogue \cite{KLRY}
of the Sethi-Vafa-Witten formula \cite{SVW}
for elliptically fibered Calabi-Yau threefolds with various fiber types.
We note that the Sethi-Vafa-Witten formula was originally
derived for elliptically fibered Calabi-Yau fourfolds in Weierstrass form,
and has been extended to more general elliptically fibered
fourfolds \cite{KLRY,AluffiEsole,Fullwood}.
While the equations we have derived are aesthetically pleasing,
we do not yet have much insight into how much they add to
what we already know about the geometry of Calabi-Yau threefolds.
Understanding the origin and implications of these equations geometrically
and possibly generalizing them in a meaningful way would be an
interesting direction of inquiry.

For the rest of the section, we discuss our results in the context
of the string universality conjecture in six dimensions \cite{SUniversality},
and conclude with some comments on
four-dimensional F-theory backgrounds.
The anomaly constraints of six dimensional supergravity
theories with minimal supersymmetry are strong enough
that it seems plausible that any ``consistent" low energy
theory---characterized by its massless spectrum---can
be realizable in string vacua.
The word consistent is in quotation marks because it is not clear
at the moment what the complete set of consistency
conditions on these theories should be.
Since, however, anomaly constraints are a necessary condition
of consistency, the strong anomaly constraints render the
space of potentially consistent theories quite manageable
under certain assumptions.\footnote{We note that
in ten dimensions, the anomaly constraints
are strong enough to impose string universality on their own
\cite{ganomalies, GS, heterotic, ADT}.}

For example, it has been proven that the number of
possible non-anomalous massless spectra is bounded
when the number of tensor multiplets $T$
is smaller than nine and there are no abelian gauge group
factors \cite{KMT2, KTBounds}. The situation becomes less
tractable when the gauge group has abelian factors.
It has been shown that while the number of allowed
gauge groups and non-abelian matter representations
are bounded when $T<9$, there exist
infinite classes of theories generated by assigning
different $U(1)$ charges to the matter \cite{PT}.

The immediate question that arises in this context is
whether such infinite classes of theories are consistent. 
This is a hard question to answer. 
To our knowledge, there are no consistency conditions
that could rule out the simple
examples of infinite classes of theories given in \cite{PT},
but at the same time, there is no guarantee that
these examples are consistent.
We may be less ambitious and ask whether there is an
obstruction to embedding all of these theories
in string theory.
This question is still a difficult one to answer,
as there may be undiscovered string vacua
with minimal supersymmetry in six dimensions.
The practical strategy to pursue seems to be
to ask whether there is an obstruction in
incorporating the infinite class of theories
to known string vacua,
and hope to gain insight from it.

An example that sheds light on this question
can be found in non-abelian theories
with no tensor multiplets \cite{KPT}.
When $T=0$, all the anomaly equations
simplify as all the anomaly coefficients
become numbers.
Due to this fact, there is a systematic
way of constructing all non-anomalous models
given the gauge group.
Hence it is possible to compare all non-anomalous
models with all known string vacua for
a given gauge group.

\begin{table}[!t]
\centering
\begin{tabular}{|c|c|c|c|c|}
\hline
N &   \# $SU(N)$ models & \# satisfy Kodaira\\
\hline
13-24 & 1 & 1\\
12 & 2 & 2\\
11 & 2 & 2\\
10 & 2 & 2\\
9 & 3 & 3\\
8 & 15 & 14\\
7 & 16 & 16\\
6 & 48 & 47\\
5 & 23 & 16\\
4 & 207 & 154\\
3 & 10100 & 262 \\
2 & $\sim 5\times 10^{7}$ & 176\\
\hline
\end{tabular}
\caption{The table of numbers of non-anomalous
  $T=0$ theories with gauge group $SU(N)$ for various $N$.
  There are no non-anomalous theories when $N>24$.
  The number of all theories that are anomalous are given in
  the second column. The number of theories that satisfy the Kodaira
  condition in addition are given in the third column.
  The number of non-anomalous theories with $N=2$ are very
  large and have not been computed precisely.}
\label{t:nvk}
\end{table}  

All six-dimensional $(1,0)$ vacua
constructed in string theory that we are aware of satisfy
the Kodaira constraint\footnote{See,
for example, section 4 of \cite{TaylorTASI}.}
\cite{MV1,MV2}:
\be
\label{kodaira}
\sum_\kappa \nu_\kappa ( j \cdot b_\kappa) \leq -12 j \cdot a \,.
\ee
Recall that $a$, $b_\kappa$ are anomaly coefficients
from \eq{x4} and \eq{fstrength} and $j$ is the $SO(1,T)$
unit vector that parametrizes the vacuum expectation
value of the scalars of the tensor multiplet.
$\kappa$ label the non-abelian gauge groups
of the theory.
$\nu_\kappa$ are positive coefficients that are determined
by the gauge group.
The number of non-anomalous theories with gauge group $SU(N)$
and the number of such theories that also satisfy \eq{kodaira}
have been compared in \cite{KPT}.
We have reproduced the results in table \ref{t:nvk}.
From this table it is clear that
the number of allowed representations grow very
fast as the gauge group becomes small.
The obstruction to embedding the bulk of the theories
in string theory---in a way known to us---is the Kodaira constraint.

In light of this example, generalizing the Kodaira
constraint in known string vacua
to incorporate abelian gauge groups,
if possible, would be an important step in addressing
the problem of the infinite classes of
theories with different $U(1)$ charge assignments.
Since $U(1)$ charges of quantum gravity theories
are expected to be quantized
\cite{Polchinski,BanksSeiberg,HellermanSharpe},
the anomaly equations \eq{gauge} imply that
the magnitude of some $b_{ii}$ must be
unbounded for any
infinite class of $U(1)$ charge assignments.
If a generalized version of the Kodaira constraint
bounds the magnitude of $b_{ii}$
at least for known string vacua,
only a finite number among the infinite non-anomalous
$U(1)$ charge assignments would
be realizable in known string vacua.

From this point of view, the results
of this paper is a step toward generalizing the
Kodaira constraint for F-theory vacua.
The geometry of the non-abelian sector
and the Kodaira constraint is well understood in F-theory.
We have used M-theory/F-theory duality to understand
the geometry of the abelian sector.

We have shown that there is a correspondence between
abelian vector fields of a six-dimensional F-theory background
obtained by compactifying on Calabi-Yau manifold $X$,
and certain four-cycles in the resolved manifold $\Xh$.
To be more precise, there is a one-to-one correspondence
between six-dimensional abelian gauge fields $A_i$
and four-cycles $S_i$ of type S in $\Xh$.
Type S four-cycles are defined to be images
of the generators of the rational sections under
the Shioda map \eq{Shioda}.
The hypermultiplets charged under the
abelian gauge fields come from quantizing zero modes of
isolated rational curves $c_r$ that shrink in the fibration limit $X \ra \Xh$.
Their charges are given by $c_r \cdot S_i$.
We have shown that the abelian anomaly coefficients
$b_{ij}$, as defined in \eq{x4} and \eq{fstrength},
are given by
\be
b_{ij} = -\pi(S_i \cdot S_j) \,.
\ee
$\pi$ is the projection to the base manifold $\BB$.
$S_i \cdot S_j$ is a curve and hence its projection
is a linear combination of two, one and zero-cycles in the base.
Treating one-cycles and zero-cycles as null vectors
in $H_2(\BB)$, we find that 
$b_{ij}$ are vectors in the $H_2(\BB)$
lattice just as the $b_\kappa$.
Whether a generalized version of the Kodaira constraint
involving the $b_{ij}$ could be derived geometrically
remains to be seen.

Four dimensional F-theory backgrounds have much
richer structure than six-dimensional backgrounds.\footnote{An
incomplete list of references on the structure of
four-dimensional F-theory backgrounds is given in the
bibliography \cite{Grimm,MSS,GKPW,Denef,DW,BHV,HayashiKawanoTatarWatari,
BlumenhagenF,CGH,BraunCollinucciValandro}.
A nice review of this subject and further references
can be found in \cite{Weigand}.}
Therefore understanding the interaction between anomaly constraints
and consistency conditions that the geometry and various fluxes of 4D
F-theory constructions must satisfy is expected to be more involved.
There has, however, been beautiful work \cite{Marsano} in which
constraints on ``hypercharge fluxes" on F-theory
$SU(5)$ GUT models with $U(1)$ symmetries---referred to as
``generalized Dudas-Palti relations" \cite{DudasPalti}---are derived by
four-dimensional anomaly cancellation.
The generalized Dudas-Palti relations provide a good handle
on F-theory GUT models with $U(1)$ symmetries \cite{Marsano,DMSS}.
It would be interesting to expand the anomaly analysis to more
general F-theory constructions and see if one could
understand the constraints that anomaly cancellation imposes upon
the various building-blocks of 4D F-theory models in the language of
intersection theory.

%It is interesting that it is claimed that
%any hyperflux distribution that satisfy the generalized Dudas-Palti
%equations can be allowed in principle in these works.
%This is certainly not the case with six-dimensional theories;
%F-theory cannot incorporate all non-anomalous models.

\section*{Acknowledgement}

First and foremost I would like to thank Wati Taylor
for his guidance, support and encouragement throughout the
process of writing this paper.
I would also like to thank Allan Adams, Koushik Balasubramanian,
Federico Bonetti, Frederik Denef, Jacques Distler, Mboyo Esole,
I\~naki Garc\'ia-Etxebarria, James Fullwood, Thomas Grimm, James Halverson,
Abhinav Kumar, Vijay Kumar, Joseph Marsano,  
David Morrison, Wati Taylor and Martijn Wijnholt for valuable discussions.
I would especially like to thank Federico Bonetti and
Thomas Grimm for sharing unpublished results.
I would like to thank the math and physics departments
of the University of Pennsylvania,
the Perimeter Institute for Theoretical Physics,
and the physics department of the
University of Wisconsin Madison,
the organizers of String-Math 2011,
the organizers of Holographic Cosmology 2.0,
the organizers of Fundamental Issues in Cosmology,
and the organizers of String Phenomenology 2011
for their hospitality at various stages of this work.
This work is supported in part by funds provided by the
DOE under contract \#DE-FC02-94ER40818.
I also acknowledge support as a String Vacuum Project Graduate Fellow,
funded through NSF grant PHY/0917807.

%--------------- APPENDIX ---------------------------
\appendix

\section{Lie Algebra and Intersection Theory} \label{ap:lieint}

In this appendix, we show that the normalized
coroot inner-product matrix, and the Cartan matrix defined as
\begin{align}
\CC_{IJ} &={1 \ov \lambda(\GG)}
{4 \vev{\alpha_I,\alpha_J} \ov \vev{\alpha_I,\alpha_I} \vev{\alpha_J,\alpha_J}} \\
C_{IJ} &={2 \vev{\alpha_I,\alpha_J} \ov \vev{\alpha_I,\alpha_I} }
\end{align}
are related to the intersection matrix of cycles obtained by
blowing up singular fibers.
$\alpha_I$ are simple roots of the Lie algebra.
The simple roots are normalized by fixing the normalization of
the matrices $\{ T_i \}$ that generate the Cartan sub-algebra such that
\be
\tr T_i T_j = \delta_{ij} \,,
\ee
where the trace is taken in the fundamental representation.
Therefore we see that the normalization of the roots depend on the
Lie group, rather than the Lie algebra. The Cartan matrix,
however, clearly only depends on the Lie algebra rather than the
Lie group from its definition; it is independent of the normalization
of the roots. We show that the same holds for the normalized
coroot matrix $\CC_{IJ}$ later in this section.

To make a precise statement relating these matrices to the intersection
theory of a resolved codimension-one singularity on the base,
let us set up the context.
Suppose there is a singular fiber of Calabi-Yau threefold $X$
fibered over a curve $b$ in the base that gives an enhanced
gauge symmetry with Lie algebra $\mathfrak{g}$.
One can resolve this singularity by blowing up $r$ independent
$\mathbb{P}^1$'s, where $r$ is the rank of $\mathfrak{g}$.
Denote the $r$ $\mathbb{P}^1$'s as $\chi_1, \cdots, \chi_r$.
Also denote the $r$ four-cycles obtained by fibering the $\mathbb{P}^1$'s
along $b$ as $C_1, \cdots, C_r$.
In the case of non-simply laced gauge groups,
a single fiber of $C_I$ might contain multiple copies $\mathbb{P}^1$'s
because monodromy of the fibers will map rational curves in
the fiber into each other.
We denote the monodromy invariant fibers $\gamma_I$, so that
$C_I$ is obtained by fibering $\gamma_I$ over $b$.

The statement is that
\begin{align}
\begin{split}
B_\alpha \cdot C_I \cdot C_J &=-\CC_{IJ} b_\alpha \\
C_I \cdot c_J  &= -C_{IJ}
\end{split}
\label{intcart}
\end{align}
where $B_\alpha$ are the four-cycles that are obtained by fibering
the elliptic fiber over elements of $H_2(\BB)$.
We check this statement in two steps.
In section \ref{sap:lie} we review some basic facts on Lie groups.
In particular, we list some useful properties of the coroot basis of the Cartan
sub-algebra and compute $\CC$ for all the simple Lie groups.
In section \ref{sap:int} we verify the formulae \eq{intcart}.

All of the facts stated in this appendix either can be found in,
or are implicit in \cite{MV1,MV2,AKM,GM2000,GM2011,BIKMSV,AspinwallK3},
but we have stated them in a way that is convenient
for our purposes.

\subsection{Some Lie Algebra} \label{sap:lie}

In this section, we review some relevant Lie algebra.
Almost all of what is discussed in this section can be found in standard
texts such as \cite{Georgi, Cahn}.

For a given Lie group $\GG$ and its Lie algebra $\mathfrak{g}$,
let us define the generators of Cartan sub-algebra $\{ T_i \}$.
Let us normalize the Cartan generators so that
\be
\tr T_i T_j = \delta_{ij}
\ee
where the trace is taken in the fundamental representation.
We can diagonalize all the other generators of the Lie group
with respect to $\{ T_i \}$.
Each such generator is uniquely labelled by its eigenvalue
under $\{ T_i \}$, {\it i.e.,}
\be
[T_i, E_\alpha] = \alpha_i E_\alpha \,.
\ee
In other words, there is a one-to-one correspondence between
the vectors $\alpha$ and the generators of the Lie group.
These vectors $\alpha$ are called the roots of the Lie algebra.

Notice that $\alpha$ will scale with a change of normalization in $T_i$.
Since we have normalized the Cartan generators
in an unambiguous way with respect to the definition of $\GG$,
the normalization of $\alpha$ are also fixed.
This is because the weights $\beta_s$ of the fundamental representation of
$\GG$ must satisfy
\be
\sum_s \beta_{s,i}^2 =1
\ee
for each $i$, where $\beta_{s,i}$ is the $i$ coordinate value of $\beta_{s}$.
This condition fixes the normalization of the weight lattice.
In this sense, we can say that $\alpha$ are the roots of the Lie group $\GG$,
with a slight abuse of terminology.
 
Now let us determine $\lambda(\GG)$ with respect to these vectors.
Recall that $\lambda(\GG)$ is a normalization factor fixed
by demanding that the smallest topological charge of an embedded
$SU(2)$ instanton is 1. This definition can be rephrased in the following way.

For any given Lie group $\GG$, we may find an $SU(2)$
subgroup. Hence we may always find an $SU(2)$ sub-algebra $\mathfrak{s}$
generated by a subgroup of the generators of the Lie algebra $\mathfrak{g}$ of $\GG$,
{\it i.e.,} there exist elements $S_i$, $i=1,2,3$ of the Lie algebra that satisfy
\be
[S_i,S_j] = i\epsilon_{ijk} S_k \,.
\ee
From this relation, one can deduce that
\be
2\tr S_1^2 = 2\tr S_2^2 = 2\tr S_3^2
\ee
in any representation.
Let us call this value $l(\mathfrak{s})$
where the trace is taken in the fundamental representation.
The normalization of the $S_i$ are fixed; if we multiply them by a
factor, the defining commutation relation does not hold anymore.
Therefore, for all the $SU(2)$ sub-algebras $\mathfrak{s}$ of 
Lie algebra $\mathfrak{g}$, the $l(\mathfrak{s})$ is
a well-defined number. We define $\lambda(\GG)$ to be,
\be
\lambda(\GG) =
\mathop{\text{min}}_{\{ \mathfrak{s} \} }
l(\mathfrak{s})
\ee
where $\{ \mathfrak{s} \}$ are all the $SU(2)$ sub-algebras of $\mathfrak{g}$.
For example, in $SU(2)$ the generators that satisfy the $SU(2)$
sub-algebra---in the fundamental representation---are
given by $S_i= {1 \ov 2} \sigma_i$ where $\sigma_i$
are the Pauli matrices. It is clear that $2 \tr S_1^2 =1$.
For $SO(4)$, the generators that satisfy the $SU(2)$ sub-algebra 
with minimum $l(\mathfrak{s})$ are,
\be
S_1 =
\begin{pmatrix}
0&i/2&0&0 \\ -i/2&0&0&0 \\ 0&0&0&-i/2 \\ 0&0&i/2&0
\end{pmatrix},~
S_2 =
\begin{pmatrix}
0&0&1/2&0 \\ 0&0&0&1/2 \\ 1/2&0&0&0 \\ 0&1/2&0&0
\end{pmatrix},~
S_3 =
\begin{pmatrix}
0&0&0&i/2 \\ 0&0&-i/2&0 \\ 0&-i/2&0&0 \\ i/2&0&0&0
\end{pmatrix}
\ee
In this case, $l(\mathfrak{s})=2 \tr S_1^2 = 2$.

Now it can be shown that for any root $\alpha$
\be
[{\alpha\cdot T \ov \vev{\alpha,\alpha}},E_{\alpha}] = E_{\alpha},
\quad
[{\alpha\cdot T \ov \vev{\alpha,\alpha}},E_{-\alpha}] = - E_{\alpha},
\quad
[E_{\alpha},E_{-\alpha}] \propto \alpha\cdot T\,,
\label{adott}
\ee
where we have defined
\be
\alpha\cdot T \equiv \alpha_i T_i \,.
\ee
We may use the freedom to rescale $E_\alpha$
so that the proportionality constant in \eq{adott} is ${\vev{\alpha,\alpha}^{-1}}$.
Then
\be
{E_\alpha+ E_{-\alpha} \ov 2}, \quad {E_\alpha- E_{-\alpha} \ov 2i},
\quad {\alpha \cdot T \ov \vev{\alpha,\alpha}}
\ee
generate a $\mathfrak{su}(2)$ subalgebra $\mathfrak{s}(\alpha)$ of $\mathfrak{g}$.
Then
\be
l(\mathfrak{s}(\alpha)) 
= 2 \tr ({\alpha \cdot T \ov \vev{\alpha,\alpha}})^2 
= {2 \ov \vev{\alpha,\alpha}}\,.
\ee
Every $\mathfrak{su}(2)$ sub-algebra can be embedded into the Lie algebra
in this way by a change of basis, so we find that
\be
\lambda(\GG) = {2 \ov\vev{\alpha,\alpha}_\text{max}} \label{lambroot}
\ee
where $\vev{\alpha,\alpha}_\text{max}$ is the length squared
of the longest root of the Lie algebra.

Now let us examine properties of $\{ \sT_I \}$,
which are the coroot basis for the Cartan generators.
They are defined to be
\be
\sT_I \equiv {2\alpha_{I} \cdot T \ov \vev{\alpha_I,\alpha_I}}
\ee
where $\alpha_I$ are the simple roots of the Lie group.
The charges of the root vectors $E_\beta$ under $\sT_I$ are
given as
\be
[\sT_I,E_\beta] = {2\alpha_{I,i} \ov \vev{\alpha_I,\alpha_I}}
[T_i, E_\beta] =  {2\vev{\alpha_{I},\beta} \ov \vev{\alpha_I,\alpha_I}}\,.
\ee
In particular,
\be
[\sT_I,E_{\alpha_J}] = 
{2\vev{\alpha_{I},\alpha_J} \ov \vev{\alpha_I,\alpha_I}} =C_{IJ}\,.
\ee

Now let us examine
\be
{1 \ov \lambda(\GG)} \tr \sT_I \sT_J \,.
\ee
Using \eq{lambroot} we find that
\be
{1 \ov \lambda(\GG)} \tr \sT_I \sT_J =
{2 \vev{\alpha,\alpha}_\text{max} \vev{\alpha_I,\alpha_J} \ov
\vev{\alpha_I,\alpha_I} \vev{\alpha_J,\alpha_J}} = \CC_{IJ}\,.
\ee
This is exactly the inner-product matrix for the
coroot lattice normalized such that the shortest
coroot has length $2$.
Note that although we had to refer to the group $\GG$
in defining $\sT_I$, the matrix $\CC_{IJ}$ only depends on
the Lie algebra due to the dividing out by $\lambda(\GG)$.
For example, $\CC=(2)$ for both $SU(2)$ and $SO(3)$.

For simply laced groups, $\CC_{IJ}$ coincides with
the Cartan matrix $C_{IJ}$. For non-simply laced groups
$\CC$ and $C$ are different.
$\CC$ for $B_n$ and $C_n$ are given by
\be
\CC(B_n) =
\begin{pmatrix}
2&-1&\cdots&0&0&0&0\\
-1&2&\cdots&0&0&0&0\\
\vdots&\vdots&\ddots&\vdots&\vdots&\vdots&\vdots\\
0&0&\cdots&2&-1&0&0\\
0&0&\cdots&-1&2&-1&0\\
0&0&\cdots&0&-1&2&-2\\
0&0&\cdots&0&0&-2&4
\end{pmatrix}
,\quad
\CC(C_n) =
\begin{pmatrix}
2&-2&0&0&\cdots&0&0\\
-2&4&-2&0&\cdots&0&0\\
0&-2&4&-2&\cdots&0&0\\
0&0&-2&4&\cdots&0&0\\
\vdots&\vdots&\vdots&\vdots&\ddots&\vdots&\vdots\\
0&0&0&0&\cdots&4&-2\\
0&0&0&0&\cdots&-2&4
\end{pmatrix}
\ee
For $B_n$ we have defined $\alpha_n$ to be the simple
root with the different(short) norm, {\it i.e.,}
\be
\vev{\alpha_1,\alpha_1}=
\vev{\alpha_2,\alpha_2}=
\cdots=
\vev{\alpha_{n-1},\alpha_{n-1}}=
2\vev{\alpha_n,\alpha_n} \,.
\ee
For $C_n$ we have defined $\alpha_1$ to be the simple
root with the different(long) norm, {\it i.e.,}
\be
\vev{\alpha_1,\alpha_1}=
2\vev{\alpha_2,\alpha_2}=
\cdots=
2\vev{\alpha_{n-1},\alpha_{n-1}}=
2\vev{\alpha_n,\alpha_n} \,.
\ee
Note that the coroot corresponding to a long/short root
becomes a short/long coroot, respectively.

$\CC$ for $F_4$ and $G_2$ are given by
\be
\CC(F_4) =
\begin{pmatrix}
2&-1&0&0\\
-1&2&-2&0\\
0&-2&4&-2\\
0&0&-2&4\\
\end{pmatrix}
,\quad
\CC(G_2) =
\begin{pmatrix}
2&-3\\
-3&6
\end{pmatrix}
\ee
respectively. For $F_4$ we have taken $\alpha_1$ and $\alpha_2$
to be the long roots, {\it i.e.,}
\be
\vev{\alpha_1,\alpha_1}=
\vev{\alpha_2,\alpha_2}=
2\vev{\alpha_3,\alpha_3}=
2\vev{\alpha_4,\alpha_4} \,.
\ee
For $G_2$ we have taken $\alpha_1$ to be the
long root, {\it i.e.,}
\be
\vev{\alpha_1,\alpha_1}=
3\vev{\alpha_2,\alpha_2} \,.
\ee
For each non-simply laced group, we have aligned the roots
so that they are decreasing in norm.

\subsection{Matching with Intersection Theory} \label{sap:int}

We verify the equations \eq{intcart} in this section first for
simply laced Lie algebras, and then for non-simply laced Lie algebras.
We denote the curve in the base that the blown up
singular fiber is fibered over in the resolved manifold, $b$.

We verify the equations pictorially.
For each gauge algebra, we draw the corresponding
tree of resolved rational curves and label the linearly independent
curves as $\alpha_I$ and label the monodromy
invariant combinations of rational curves $\gamma_J$ according to \cite{GM2000}.
The curves $\chi_I$ that
M2 branes wrap to give root vectors are identified with $\alpha_I$.
The four cycles $C_J$ dual to non-abelian gauge field components
of the coroot basis elements $\sT_J$ of the Cartan are obtained
by fibering $\gamma_J$ over $b$.

We verify that
\begin{align}
\begin{split}
\gamma_I \cdot \gamma_J = -\CC_{IJ} \\
\gamma_I \cdot \alpha_J = -C_{IJ}
\end{split}
\label{intcond}
\end{align}
where the intersections are taken within a local
complex dimension two slice of the manifold transverse
to $b$ at a generic point in $b$.
These two equations imply \eq{intcart} since
\begin{align}
B_\alpha \cdot C_I \cdot C_J
&= b_\alpha (\gamma_I \cdot C_J)
= -b_\alpha \CC_{IJ} \,, \\
C_I \cdot \chi_J
&= C_I \cdot \alpha_J = -C_{IJ} \,.
\end{align}
The latter equalities of the two equations
follow from \eq{intcond}
since $C_I$ is a $\gamma_I$ fibration over $b$.
We note that all the data are defined for the Lie algebra,
and not sensitive to the Lie group.

\subsubsection{Simply Laced Lie Algebras}

For simply laced Lie algebras, the monodromy group of
the blown-up singular fibers
are trivial and the blown-up rational curves
form the Dynkin diagram of the corresponding Lie algebra,
except possibly for the case of $\mathfrak{su}(2)$.
It turns out that $\alpha_I=\gamma_I$ for all the simply
laced Lie algebras.

The self-intersection number of a rational curve is $(-2)$
and the intersection number between adjacent rational curves
is $1$.
The intersection number between non-adjacent curves are $0$.
The intersection number satisfies linearity conditions, {\it i.e.,}
\be
c \cdot (\lambda_1 c_1 + \lambda_2 c_2) =
\lambda_1 (c \cdot  c_1) +\lambda_2 (c \cdot c_2) \,.
\ee
Based on these rules, we can verify that the resolved fibers that
give the $A, D, E$
algebra satisfy \eq{intcond}.

\begin{figure}[!h]
\centering\includegraphics[width=6cm]{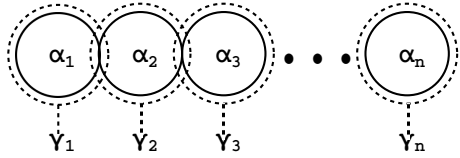}
\caption{\small Resolved fiber for $A_n$. The curves $\alpha$
corresponding to root vectors are in solid lines 
while the monodromy invariant fibers
$\gamma$ corresponding to coroots are in dotted lines.}
\label{f:A}
\end{figure}

$I_{n+1}$ fibers, or possibly the $III$/$IV$ fiber for $A_1$/$A_2$
respectively, give the $A_n$ Lie algebra.
The tree of blown-up rational
curves of the resolved $I_{n+1}$ fiber---or
the $III$/$IV$ fiber for $A_1$/$A_2$---is
depicted in figure \ref{f:A}.
It is clear that
\be
(\gamma_I \cdot \gamma_J)
=(\gamma_I \cdot \alpha_J)
=
\begin{pmatrix}
-2&1&\cdots&0&0&0&0\\
1&-2&\cdots&0&0&0&0\\
\vdots&\vdots&\ddots&\vdots&\vdots&\vdots&\vdots\\
0&0&\cdots&-2&1&0&0\\
0&0&\cdots&1&-2&1&0\\
0&0&\cdots&0&1&-2&1\\
0&0&\cdots&0&0&1&-2
\end{pmatrix}
=-C_{IJ}=-\CC_{IJ} \,.
\ee

\begin{figure}[!h]
\centering\includegraphics[width=6cm]{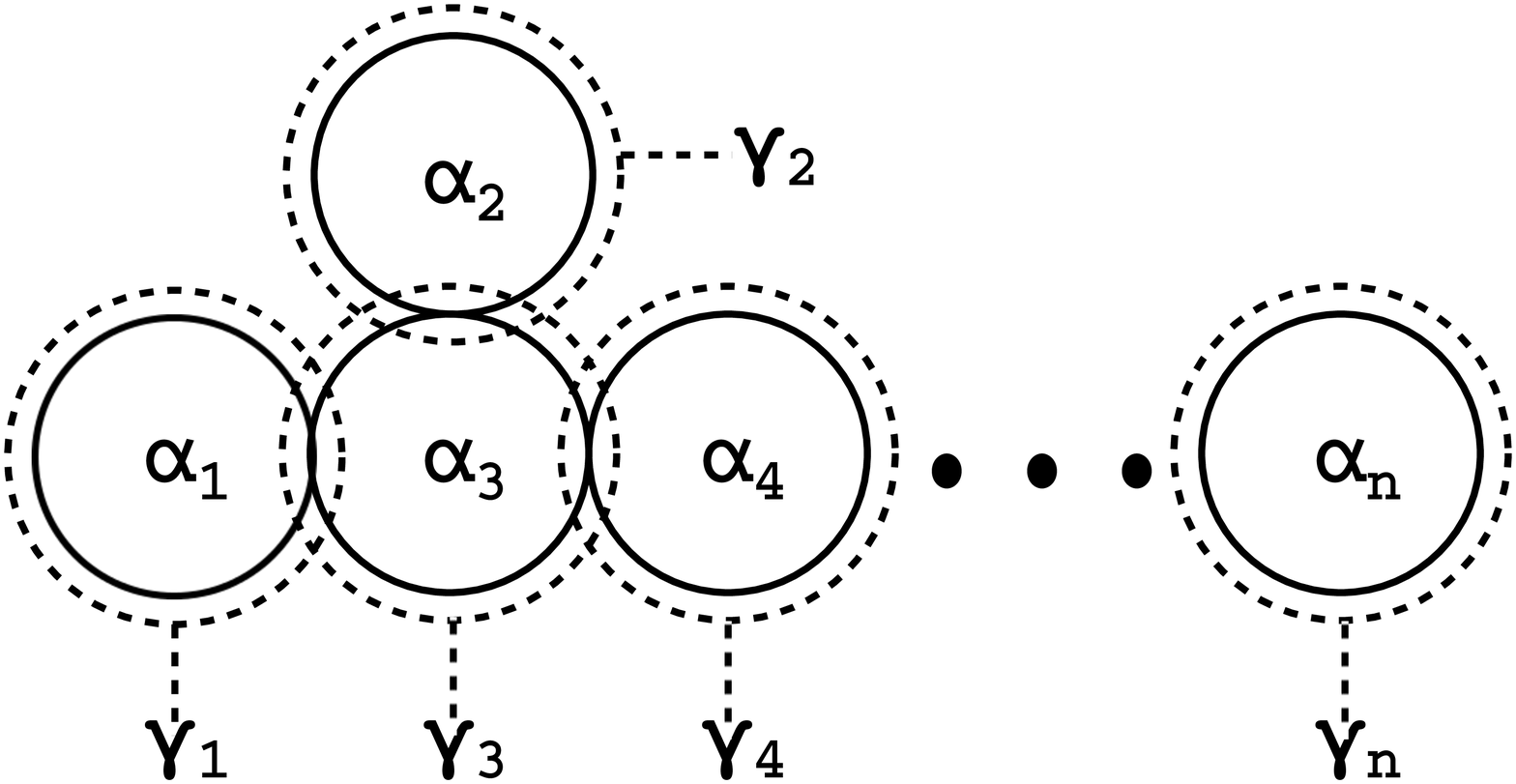}
\caption{\small Resolved fiber for $D_n$.}
\label{f:D}
\end{figure}

$I^*_{n-4}$ fibers give the $D_n$ Lie algebra.
The tree of blown-up rational
curves of the resolved $I^*_{n-4}$ fiber is
depicted in figure \ref{f:D}.
The intersection matrices are given by
\be
(\gamma_I \cdot \gamma_J)
=(\gamma_I \cdot \alpha_J)
=
\begin{pmatrix}
-2&0&1&0&\cdots&0&0\\
0&-2&1&0&\cdots&0&0\\
1&1&-2&1&\cdots&0&0\\
0&0&1&-2&\cdots&0&0\\
\vdots&\vdots&\vdots&\vdots&\ddots&\vdots&\vdots\\
0&0&0&0&\cdots&-2&1\\
0&0&0&0&\cdots&1&-2
\end{pmatrix}
=-C_{IJ}=-\CC_{IJ} \,.
\ee

\begin{figure}[!h]
\centering\includegraphics[width=6cm]{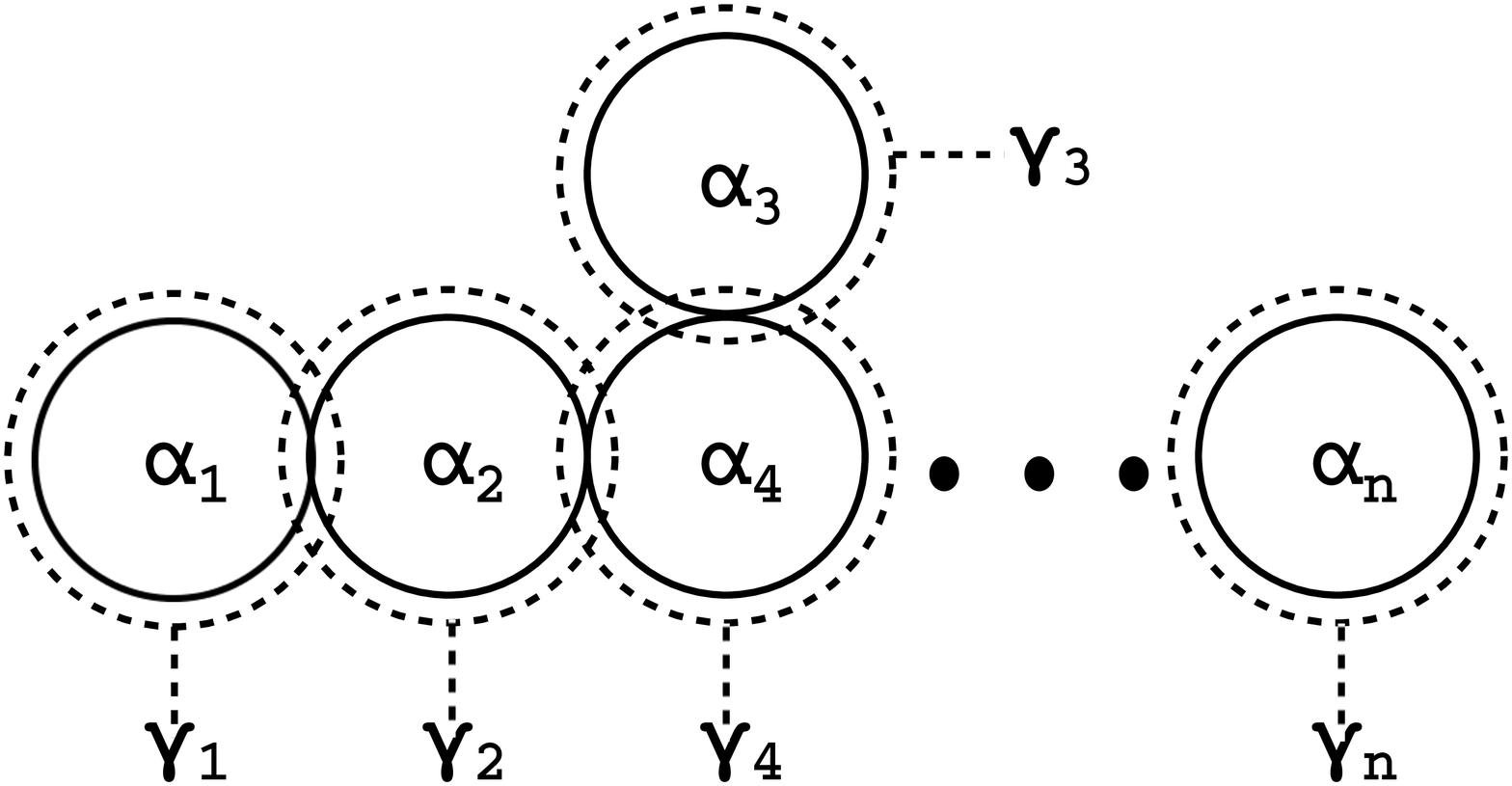}
\caption{\small Resolved fiber for $E_n$.}
\label{f:E}
\end{figure}

The fibers $IV^*$, $III^*$ and $II^*$
give the $E_6$, $E_7$ and $E_8$ Lie algebra respectively.
The tree of blown-up rational
curves of the resolved $E_n$ fiber is
depicted in figure \ref{f:E}.
The intersection matrices are given by
\be
(\gamma_I \cdot \gamma_J)
=(\gamma_I \cdot \alpha_J)
=
\begin{pmatrix}
-2&1&0&0&0&\cdots&0&0\\
1&-2&0&1&0&\cdots&0&0\\
0&0&-2&1&0&\cdots&0&0\\
0&1&1&-2&1&\cdots&0&0\\
0&0&0&1&-2&\cdots&0&0\\
\vdots&\vdots&\vdots&\vdots&\vdots&\ddots&\vdots&\vdots\\
0&0&0&0&0&\cdots&-2&1\\
0&0&0&0&0&\cdots&1&-2
\end{pmatrix}
=-C_{IJ}=-\CC_{IJ} \,.
\ee

\begin{figure}[!h]
\centering\includegraphics[width=3cm]{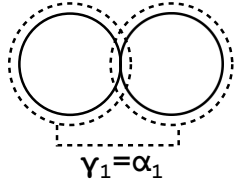}
\caption{\small Resolved fiber for $A_1$.}
\label{f:A1}
\end{figure}

$\mathfrak{su}(2)$ can come from an $I_3$ or $IV$
fiber with a $\field{Z}_2$ monodromy.
The resolved fiber that gives the $A_1$
Lie algebra in this way is given by figure \ref{f:A1}.
The $\field{Z}_2$ interchanges the two $\field{P}^1$'s
drawn in the figure, and hence
the point where the two rational curves touch
is singular.
$\gamma_1$ is the monodromy invariant fiber.
It is shown in \cite{AKM} that the BPS states come
from branes wrapping $\alpha_1$ rather than the
the individual components drawn as spheres in
the figure.
The intersection matrices are given by
\be
(\gamma_I \cdot \gamma_J)
=(\gamma_I \cdot \alpha_J)
=
\begin{pmatrix}
-2
\end{pmatrix}
=-C_{IJ}=-\CC_{IJ} \,.
\ee

\subsubsection{Non-simply Laced Lie Algebras}

For non-simply laced Lie algebras, the blown-up singular fibers
have non-trivial monodromy. The blown-up rational curves
form the Dynkin diagram of a larger Lie algebra.
Under monodromy, the rational curves are exchanged
among themselves.
For each fiber, we denote the independent rational curves $\alpha_I$,
and the monodromy invariant components of the fiber $\gamma_I$ .
Let us verify that the resolved fibers that give the $A, D, E$
algebra satisfy \eq{intcond}.

\begin{figure}[!h]
\centering\includegraphics[width=6cm]{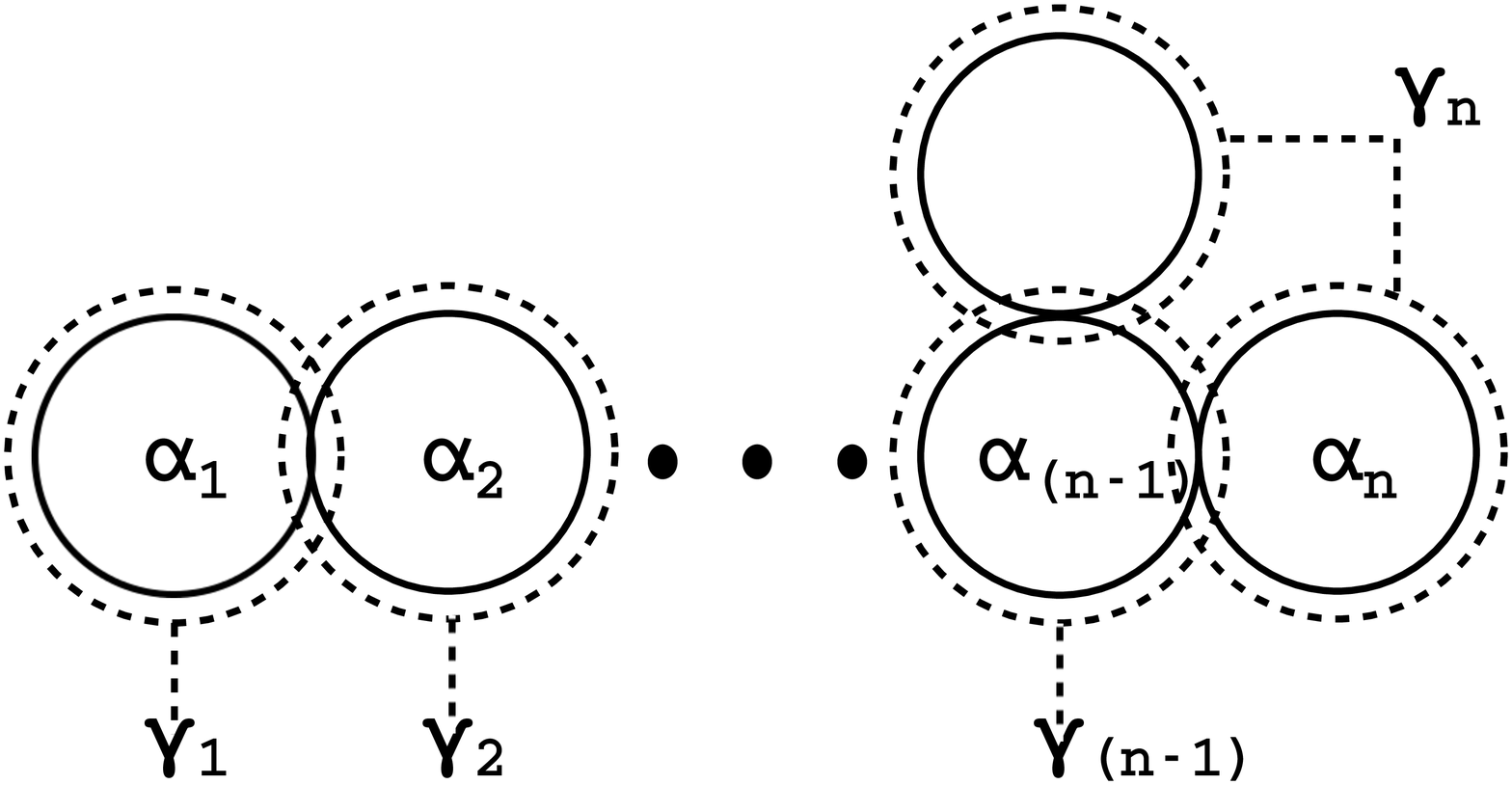}
\caption{\small Resolved fiber for $B_n$. The curves $\alpha$
corresponding to root vectors are in solid lines 
while the monodromy invariant fibers
$\gamma$ corresponding to coroots are in dotted lines.}
\label{f:B}
\end{figure}

The fibers $I^*_{(n-3)}$ with $\field{Z}_2$ monodromy
give the $B_n$ Lie algebra.
The tree of blown-up rational
curves of the resolved $I^*_{(n-3)}$ fiber is
depicted in figure \ref{f:B}.
The $\field{Z}_2$ monodromy exchanges the
two rational curves in $\gamma_n$.
The intersection matrices are given by
\begin{align}
\begin{split}
(\gamma_I \cdot \gamma_J)
=
\begin{pmatrix}
-2&1&\cdots&0&0&0\\
1&-2&\cdots&0&0&0\\
\vdots&\vdots&\ddots&\vdots&\vdots&\vdots\\
0&0&\cdots&-2&1&0\\
0&0&\cdots&1&-2&2\\
0&0&\cdots&0&2&-4
\end{pmatrix}
\,, ~~
(\gamma_I \cdot \alpha_J)
=
\begin{pmatrix}
-2&1&\cdots&0&0&0\\
1&-2&\cdots&0&0&0\\
\vdots&\vdots&\ddots&\vdots&\vdots&\vdots\\
0&0&\cdots&-2&1&0\\
0&0&\cdots&1&-2&1\\
0&0&\cdots&0&2&-2
\end{pmatrix}
\\[3pt] \Rightarrow
(\gamma_I \cdot \gamma_J) = -\CC_{IJ}, \quad
(\gamma_I \cdot \alpha_J) = -C_{IJ} \,.
\end{split}
\end{align}

\begin{figure}[!h]
\centering\includegraphics[width=8cm]{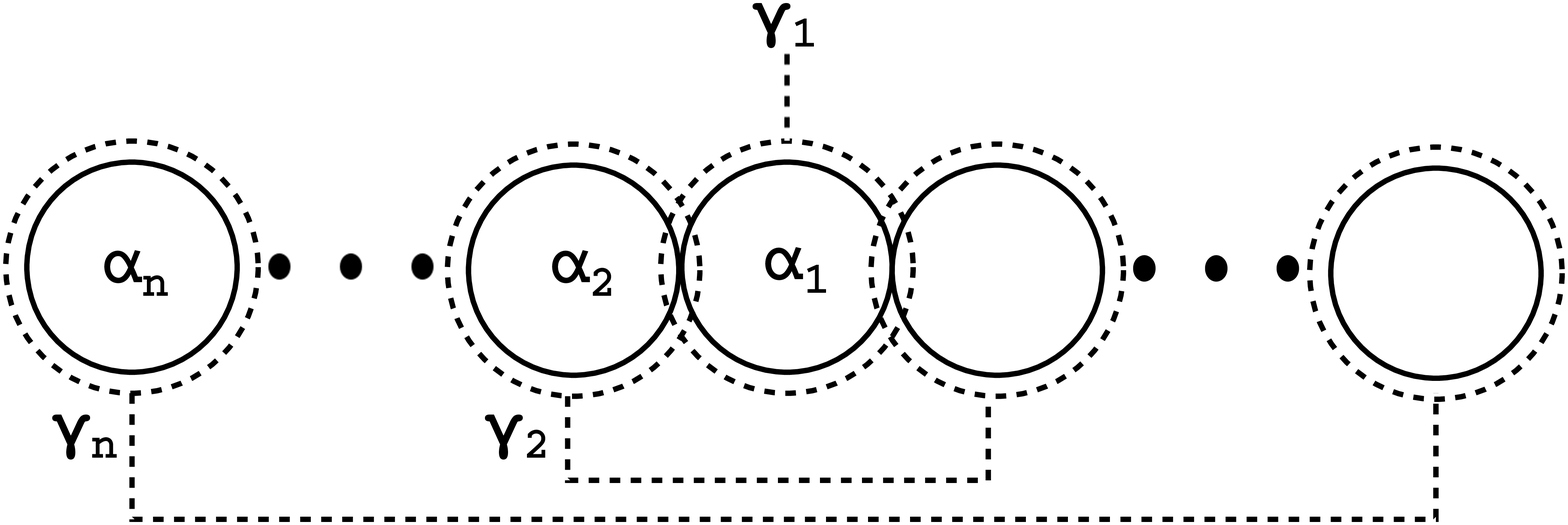}
\vskip0.3in
\centering\includegraphics[width=8cm]{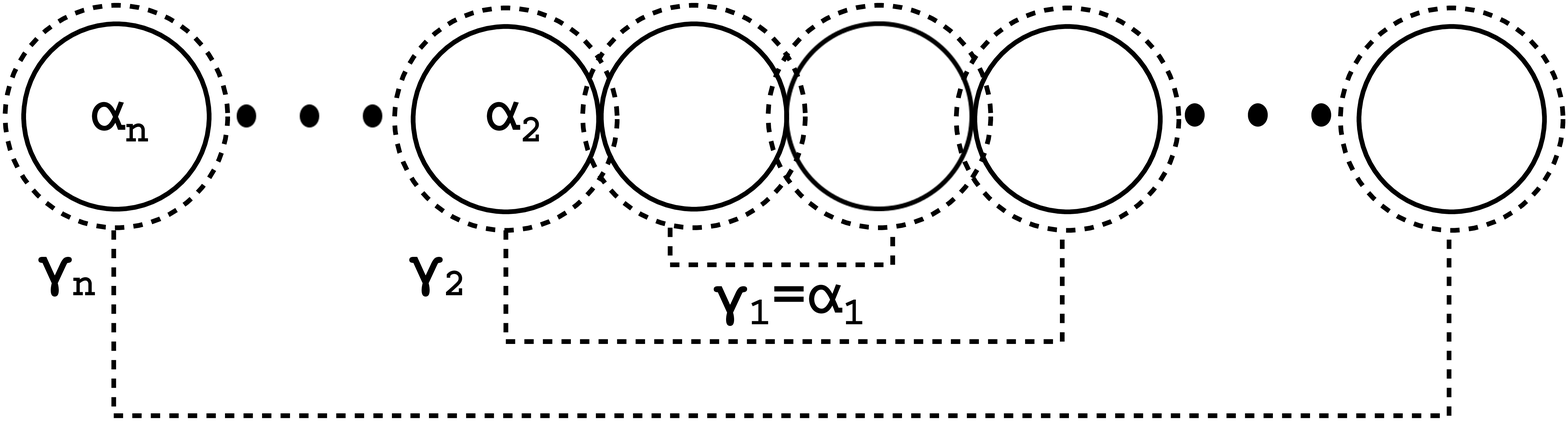}
\caption{\small Resolved fiber for $C_n$.}
\label{f:C}
\end{figure}

The fibers $I_{2n}$ or $I_{(2n+1)}$
with $\field{Z}_2$ monodromy
give the $C_n$ Lie algebra.
The trees of blown-up rational
curves of the resolved $I_{2n}$ and $I_{(2n+1)}$ fibers
are depicted in figure \ref{f:C}.
The $\field{Z}_2$ monodromy exchanges the
two rational curves in each $\gamma_I$.
Just as with the case of $\mathfrak{su}(2)$,
$\alpha_1$ should be taken to be equal to $\gamma_1$
\cite{AKM}.
The intersection matrices are given by
\begin{align}
\begin{split}
(\gamma_I \cdot \gamma_J)
=
\begin{pmatrix}
-2&2&0&\cdots&0&0\\
2&-4&2&\cdots&0&0\\
0&2&-4&\cdots&0&0\\
\vdots&\vdots&\vdots&\ddots&\vdots&\vdots\\
0&0&0&\cdots&-4&2\\
0&0&0&\cdots&2&-4
\end{pmatrix}
\,, \quad
(\gamma_I \cdot \alpha_J)
=
\begin{pmatrix}
-2&1&0&\cdots&0&0\\
2&-2&1&\cdots&0&0\\
0&1&-2&\cdots&0&0\\
\vdots&\vdots&\vdots&\ddots&\vdots&\vdots\\
0&0&0&\cdots&-2&1\\
0&0&0&\cdots&1&-2
\end{pmatrix}
\\[3pt] \Rightarrow
(\gamma_I \cdot \gamma_J) = -\CC_{IJ}, \quad
(\gamma_I \cdot \alpha_J) = -C_{IJ} \,.
\end{split}
\end{align}

\begin{figure}[!h]
\centering\includegraphics[width=6cm]{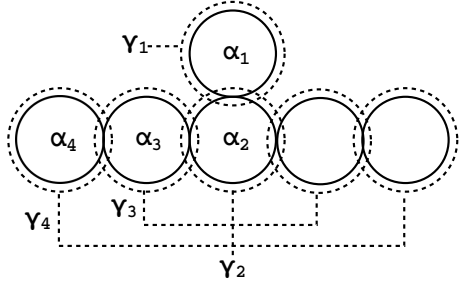}
\caption{\small Resolved fiber for $F_4$.}
\label{f:F}
\end{figure}

The fiber $IV^*$
with $\field{Z}_2$ monodromy
gives the $F_4$ Lie algebra.
The tree of blown-up rational
curves of the resolved $IV^*$ fiber
is depicted in figure \ref{f:F}.
The $\field{Z}_2$ monodromy exchanges the
two rational curves in $\gamma_3$ and $\gamma_4$.
The intersection matrices are given by
\be
(\gamma_I \cdot \gamma_J)
=
\begin{pmatrix}
-2&1&0&0\\
1&-2&2&0\\
0&2&-4&2\\
0&0&2&-4\\
\end{pmatrix}
=-\CC_{IJ} \,,~~
(\gamma_I \cdot \alpha_J)
=
\begin{pmatrix}
-2&1&0&0\\
1&-2&1&0\\
0&2&-2&1\\
0&0&1&-2\\
\end{pmatrix}
=-C_{IJ} \,.
\ee

\begin{figure}[!h]
\centering\includegraphics[width=4cm]{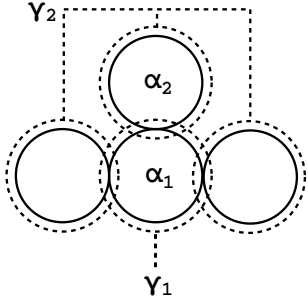}
\caption{\small Resolved fiber for $G_2$.}
\label{f:G}
\end{figure}

The fiber $I^*_0$
with $\field{Z}_3$ or $\mathfrak{S}_3$
monodromy gives the $G_2$ Lie algebra.
The tree of blown-up rational
curves of the resolved $I^*_0$ fiber
is depicted in figure \ref{f:G}.
The $\field{Z}_3$ or $\mathfrak{S}_3$
monodromy exchanges the
three rational curves in $\gamma_2$.
The intersection matrices are given by
\be
(\gamma_I \cdot \gamma_J)
=
\begin{pmatrix}
-2&3\\
3&-6
\end{pmatrix}
=-\CC_{IJ} \,,
\quad
(\gamma_I \cdot \alpha_J)
=
\begin{pmatrix}
-2&1\\
3&-2
\end{pmatrix}
=-C_{IJ} \,.
\ee

\section{Proof of Intersection Equations for $\SS_n$ of Type S or C}
\label{ap:2and3}

We prove the intersection equations
\begin{align}
\begin{split}
&\pi(\SS_1 \cdot \SS_2)\cdot \pi(\SS_3 \cdot \SS_4)+
\pi(\SS_1 \cdot \SS_3)\cdot \pi(\SS_2 \cdot \SS_4)+
\pi(\SS_1 \cdot \SS_4)\cdot \pi(\SS_2 \cdot \SS_3) \\
&=\sum_r (c_r \cdot \SS_1) (c_r \cdot \SS_2) (c_r \cdot \SS_3) (c_r \cdot \SS_4)
+ \sum_\rho (2g_\rho-2)(\chi_\rho \cdot \SS_1) (\chi_\rho \cdot \SS_2) (\chi_\rho \cdot \SS_3) (\chi_\rho \cdot \SS_4)
\end{split}
\label{ap1}
\end{align}
and
\begin{align}
\begin{split}
6K \cdot \pi(\SS_1 \cdot \SS_2) 
=\sum_r (c_r \cdot \SS_1) (c_r \cdot \SS_2) 
+ \sum_\rho (2g_\rho-2)(\chi_\rho \cdot \SS_1) (\chi_\rho \cdot \SS_2) 
\end{split}
\label{ap2}
\end{align}
when
\be
\SS_n \in \{ T_{I,\kappa}, S_i \}\,.
\ee

We first check the case when $\SS_n$ are all of
type S, then check the case when $\SS_n$ are all of
type C. Finally we check the case when there is a mixture
of type S and C cycles among $\SS_n$.
We refer to the first equation \eq{ap1} the quartic equation
and the second equation \eq{ap2} the quadratic equation
throughout this appendix.

\subsection{Type S Cycles Only}

Take $\SS_n$ to be type S cycles $\{ S_i, S_j, S_k, S_l \}$.
$\SS_n$ are dual to abelian vector multiplets.
Then using the result \eq{mainresult}
\be
b_{ij} = - \pi(S_i \cdot S_j) \,,
\ee
the last equation of
\eq{gauge} can be rewritten in the form
\begin{align}
\begin{split}
\pi(S_i \cdot S_j)\cdot \pi(S_k \cdot S_l)+
\pi(S_i \cdot S_k)\cdot \pi(S_j \cdot S_l)+
\pi(S_i \cdot S_l)\cdot \pi(S_j \cdot S_k)
=\sum_x q^x_i q^x_j q^x_k q^x_l \,.
\end{split}
\end{align}
We have used $x$ to index all the hypermultiplets in the theory
and $q^x_n$ denotes the charge of hypermultiplet $x$
under the $U(1)$ vector field dual to $S_n$.

We note that all hypermultiplets charged under
these vector multiplets come from M2 branes wrapping
type I curves, which are precisely $c_r$.
Recall that
\be
S_j \cdot \chi_\rho=0 
\ee
for all $\rho$ by the construction of S-type cycles.

Then since the charge of the hypermultiplet coming from
wrapping branes on $c_r$
under vector field $n$ is $c_r \cdot S_i$,
the last equation of \eq{gauge} is indeed equivalent to the equation
\begin{align}
\begin{split}
&\pi(S_i \cdot S_j)\cdot \pi(S_k \cdot S_l)+
\pi(S_i \cdot S_k)\cdot \pi(S_j \cdot S_l)+
\pi(S_i \cdot S_l)\cdot \pi(S_j \cdot S_k) \\
&=\sum_r (c_r \cdot S_i) (c_r \cdot S_j) (c_r \cdot S_k) (c_r \cdot S_l)
+\sum_\rho (2g_\rho -2) (\chi_\rho \cdot S_i)(\chi_\rho \cdot S_j)(\chi_\rho \cdot S_k)(\chi_\rho \cdot S_l)
\end{split}
\end{align}
since the latter term on the right hand side is zero.

Similarly, since the vector $a$ is identified with
the canonical class of the base $K$,
the second equation in \eq{mixed} implies that
\begin{align}
\begin{split}
6K \cdot \pi(S_i \cdot S_j) &= -6a \cdot b_{ij}
=\sum_x q^x_i q^x_j \\
&=\sum_r (c_r \cdot S_i)(c_r \cdot S_j)
+ \sum_\rho (2g_\rho-2) (\chi_\rho \cdot S_i)(\chi_\rho \cdot S_j) 
\end{split}
\end{align}
for two type S four-cycles $S_i, S_j$.
Hence we have shown that \eq{ap1} and \eq{ap2} hold when
$\SS_n$ are all of type S.

\subsection{Type C Cycles Only}

\noindent
We prove
\begin{align}
\begin{split}
\pi(T_{I,\kappa} \cdot T_{J,\nu})\cdot &\pi(T_{K,\lambda} \cdot T_{L,\mu})+
\text{(2 other groupings)} \\
&=\sum_r (c_r \cdot T_{I,\kappa}) (c_r \cdot T_{J,\nu}) (c_r \cdot T_{K,\lambda}) (c_r \cdot T_{L,\mu}) \\
&+ \sum_\rho (2g_\rho-2)
(\chi_\rho \cdot T_{I,\kappa}) (\chi_\rho \cdot T_{J,\nu}) (\chi_\rho \cdot T_{K,\lambda}) (\chi_\rho \cdot T_{L,\mu})
\end{split}
\label{C1}
\end{align}
and
\begin{align}
\begin{split}
6K \cdot \pi(T_{I,\kappa} \cdot T_{J,\nu})
=\sum_r (c_r \cdot T_{I,\kappa}) (c_r \cdot T_{J,\nu}) 
+ \sum_\rho (2g_\rho-2)(\chi_\rho \cdot T_{I,\kappa}) (\chi_\rho \cdot T_{J,\nu})
\end{split}
\label{C2}
\end{align}
for $T_{I,\kappa}$ of type C.
The quartic equation is only non-trivial when the $\kappa, \nu, \lambda, \mu$ are
equal in pairs---this includes the case when they are all equal.
The quadratic equation is only non-trivial when $\kappa$ and $\nu$ are equal.

The two statements above follow from three facts.
\ben
\item
$\pi(T_{I,\kappa} \cdot T_{J,\nu}) $ satisfies
\be
\pi(T_{I,\kappa} \cdot T_{J,\nu}) = -\delta_{\kappa\nu} b_\kappa \CC_{IJ,\kappa} \,,
\ee
so the left-hand side of the quartic equation is zero unless
$\kappa, \nu, \lambda, \mu$ are
equal in pairs.
Similarly, the left hand side of the quadratic equation is zero unless
$\kappa$ and $\nu$ are equal.
\item Unless $\kappa, \nu, \lambda, \mu$ are
equal in pairs,
\be
\sum_r (c_r \cdot T_{I,\kappa}) (c_r \cdot T_{J,\nu}) (c_r \cdot T_{K,\lambda}) (c_r \cdot T_{L,\mu})
= \sum_R k_R \tr_R \sT_{M,\eta} =0
\ee
for some constants $k_R$ and $M, \eta$.
Similarly, unless $\kappa$ and $\nu$ are equal,
\be
\sum_r (c_r \cdot T_{I,\kappa}) (c_r \cdot T_{J,\nu})
= \sum_R k_R \tr_R \sT_{M,\eta} =0
\ee
for some constants $k_R$ and $M, \eta$.
This is because the hypermultiplets coming from type I cycles
always can be organized into representations of the Lie algebra.
\item
$\chi_\rho$ can be organized into positive(or negative, depending
on convention) roots
of the simple Lie algebra factors as $\{ \chi_\rho \} =\{ \chi_{s,\kappa} \}$.
Any $\chi_{s,\kappa}$ for a given $\kappa$ is a linear combination of
curves $\chi_{I,\kappa}$ corresponding to the simple roots of $\GG_\kappa$.
Since
\be
\chi_{I,\kappa} \cdot T_{J,\nu} =-\delta_{\kappa\nu} C_{IJ,\kappa} \,,
\ee
the equation
\be
(\chi_\rho \cdot T_{I,\kappa})(\chi_\rho \cdot T_{I,\nu})=0
\ee
holds for $\kappa \neq \nu$.
\een
Therefore \eq{C1} is non-trivial only when $\kappa, \nu, \lambda, \mu$ are
equal in pairs, and \eq{C2} is non-trivial only when $\kappa=\nu$.

Now let us write the anomaly equations in a form more convenient
to our purposes. The anomaly equations concerning only
non-abelian gauge group
factors implies that the following holds for all elements $\tk, \tn$
of the Cartan of the gauge groups $\GG_\kappa, \GG_\nu$:
\begin{align}
a\cdot {b_\kappa } {\tr \tk^2 \ov \lambda_\kappa}&=
{1 \ov 6} (\tr_{\text{Adj}_\kappa} \tk^2 -\sum_I x_R \tr_R \tk^2)\\
b_\kappa \cdot b_\kappa  ({\tr \tk^2 \ov \lambda_\kappa})^2&=
{1 \ov 3} (\sum_I x_R \tr_R \tk^4 -\tr_{\text{Adj}_\kappa} \tk^4)\\
b_\kappa \cdot b_\nu
({\tr \tk^2  \ov \lambda_\kappa} ) ({ \tr \tn^2 \ov \lambda_\nu} ) &=
\sum_{R,S} x_{RS} \tr_R \tk^2 \tr_{S} \tn^2 \quad (\kappa \neq \nu)
\end{align}

Let us take $\tk=t_I \Tk{I}$ and $\tn=s_I \Tk{I}$ where $I$
runs over the indices of the coroot basis of the Cartan sub-algebra of each
gauge group.
Expanding the equalities above, we obtain polynomials with respect
to $t_I$ and $s_I$ on both sides of the equations. 
The anomaly equations must hold for any value of $t_I$ and $s_I$.
Hence all the coefficients of the polynomials must be identical.
By identifying the coefficients, we obtain
\begin{align}
\begin{split}
a\cdot {b_\kappa } \CC_{IJ,\kappa} &=
{1 \ov 6} (\tr_{\text{Adj}_\kappa} \Tk{I}\Tk{J} -\sum_I x_R \tr_R \Tk{I}\Tk{J})\\
b_\kappa \cdot b_\kappa
( \CC_{IJ,\kappa}  \CC_{KL,\kappa} +& \CC_{IK,\kappa}  \CC_{JL,\kappa}  
+\CC_{IL,\kappa}  \CC_{JK,\kappa}  )= \\
& \sum_I x_R \tr_R \Tk{I}\Tk{J}\Tk{K}\Tk{L} -\tr_{\text{Adj}_\kappa}  \Tk{I}\Tk{J}\Tk{K}\Tk{L}\\
b_\kappa \cdot b_\nu
\CC_{IJ,\kappa} \CC_{KL,\nu}  &=
\sum_{R,S} x_{RS} \tr_R \Tk{I} \Tk{J} \tr_S \Tn{K} \Tn{L}
\end{split}
\end{align}
for $\kappa \neq \nu$.

We can write all the elements of the right-hand sides as
a sum of products of the charge of each vector or hypermultiplet
under each Cartan element.
Each charged multiplet corresponds to a type I or a type F rational curve,
and its charges are given by the intersection numbers of the
curve with the four-cycles of type C.
Rewriting the right-hand sides of the equations we obtain
\begin{align}
\begin{split}
-6 a\cdot {b_\kappa }& \CC_{IJ,\kappa} = \\
& \sum_r (c_r \cdot T_{I,\kappa}) (c_r \cdot T_{J,\kappa})
+  \sum_\rho (2g_\rho-2) (\chi_\rho \cdot T_{I,\kappa}) (\chi_\rho \cdot T_{J,\kappa}) \\
b_\kappa \cdot b_\kappa
( \CC_{IJ,\kappa}  \CC_{KL,\kappa} &+ \CC_{IK,\kappa}  \CC_{JL,\kappa}  
+\CC_{IL,\kappa}  \CC_{JK,\kappa}  )= \\
& \sum_r (c_r \cdot T_{I,\kappa}) (c_r \cdot T_{J,\kappa})
(c_r \cdot T_{K,\kappa}) (c_r \cdot T_{L,\kappa}) \\
&+\sum_\rho (2g_\rho -2) (\chi_\rho \cdot T_{I,\kappa}) (\chi_\rho \cdot T_{J,\kappa})
(\chi_\rho \cdot T_{K,\kappa}) (\chi_\rho\cdot T_{L,\kappa})
\\
b_\kappa \cdot b_\nu
&\CC_{IJ,\kappa} \CC_{KL,\nu}  = \\
&\sum_r (c_r \cdot T_{I,\kappa}) (c_r \cdot T_{J,\kappa})
(c_r \cdot T_{K,\nu}) (c_r \cdot T_{L,\nu}) \\
&+\sum_\rho (2g_\rho -2) (\chi_\rho \cdot T_{I,\kappa}) (\chi_\rho \cdot T_{J,\kappa})
(\chi_\rho \cdot T_{K,\nu}) (\chi_\rho\cdot T_{L,\nu})
\end{split}
\end{align}
Note that the curve $\chi_\rho$ contributes
$2g_\rho$ hypermultiplets and two
vector multiplets \cite{WittenPT,KMP},
as explained in section \ref{sss:mf}.
The vector multiplet always contributes with a negative
sign with respect to the contribution of the hypermultiplet
to the right hand sides of the equations.
The last term of the last equation is zero.

Finally using
\be
\pi(T_{I,\kappa} \cdot T_{J,\nu}) = -\delta_{\kappa\nu} b_\kappa \CC_{IJ,\kappa} \,,
\ee
and the fact that $a$ is equal to the canonical class $K$ of the base,
the three equations translate into
\begin{align}
\begin{split}
6K &\cdot\pi(T_{I,\kappa} \cdot T_{J,\kappa})  = \\
& \sum_r (c_r \cdot T_{I,\kappa}) (c_r \cdot T_{J,\kappa})
+  \sum_\rho (2g_\rho-2) (\chi_\rho \cdot T_{I,\kappa}) (\chi_\rho \cdot T_{J,\kappa}) \\
\pi(T_{I,\kappa} \cdot T_{J,\kappa}) \cdot &\pi(T_{K,\kappa} \cdot T_{L,\kappa}) +
\text{(2 other groupings)} = \\
& \sum_r (c_r \cdot T_{I,\kappa}) (c_r \cdot T_{J,\kappa})
(c_r \cdot T_{K,\kappa}) (c_r \cdot T_{L,\kappa}) \\
&+\sum_\rho (2g_\rho -2) (\chi_\rho \cdot T_{I,\kappa}) (\chi_\rho \cdot T_{J,\kappa})
(\chi_\rho \cdot T_{K,\kappa}) (\chi_\rho\cdot T_{L,\kappa})
\\
\pi(T_{I,\kappa} \cdot T_{J,\kappa}) \cdot &\pi(T_{K,\nu} \cdot T_{L,\nu}) +
\text{(2 other groupings)}   = \\
&\sum_r (c_r \cdot T_{I,\kappa}) (c_r \cdot T_{J,\kappa})
(c_r \cdot T_{K,\nu}) (c_r \cdot T_{L,\nu}) \\
&+\sum_\rho (2g_\rho -2) (\chi_\rho \cdot T_{I,\kappa}) (\chi_\rho \cdot T_{J,\kappa})
(\chi_\rho \cdot T_{K,\nu}) (\chi_\rho\cdot T_{L,\nu})
\end{split}
\end{align}
We note that in the last equation, the two other groupings of cycles
that are not written down are zero.

\subsection{Both Type S and C Cycles}

The quadratic equation is always trivial when $\SS_1$ and $\SS_2$
are each of type S and C, for the following reasons.
\ben
\item The left hand side of the quadratic equation is trivially zero since
\be
\pi(S_i \cdot T_{I,\kappa}) =0 \,,
\ee
due to the construction of type S cycles.
\item The hypermultiplets come in representations of
the non-abelian Lie algebra. Hence,
\be
\sum_r (c_r \cdot S_i ) (c_r \cdot T_{I,\kappa}) = \sum_R k_R \tr_R \sT_{I,\kappa} =0\,.
\ee
\item $S_i$ do not intersect type $\chi_\rho$ curves by construction.
Hence,
\be
\sum_\rho(2g_\rho-2) (\chi_\rho \cdot S_i ) (\chi_\rho \cdot T_{I,\kappa}) = 0\,.
\ee
\een

Similarly we can show that
the non-trivial cases to
check for the quartic equation are,
without loss of generality, when
\ben
\item $\SS_1 =\Tk{I},~ \SS_2 =\Tk{J},~ \SS_3 =\Tk{K},~ \SS_4 = S_i$.
\item $\SS_1 =\Tk{I},~ \SS_2 =\Tk{J},~ \SS_3 =S_i,~ \SS_4 = S_j$.
\een
From the anomaly equations, we can show that for
any element of the Cartan $T_\kappa$ of $\GG_\kappa$
\begin{align}
\begin{split}
0 &= \sum_{R,q_i} x_{R,q_i}  q_i \tr_R T_\kappa^3  \\
b_\kappa \cdot b_{ij} {\tr T_\kappa^2 \ov \lambda_\kappa}
&= \sum_{R,q_i , q_j} x_{R,q_i,q_j}  q_i q_j \tr T_\kappa^2 
\end{split}
\end{align}
Setting $T_\kappa =\Tk{I} t_I$, we can write both sides of
the two equations as polynomials with respect to $t_I$.
Since the equality must hold for all values of $t_I$,
the coefficients of the polynomials must match on both sides,
and hence
\begin{align}
\begin{split}
0 &= \sum_{R,q_i} x_{R,q_i} \tr_R( \Tk{I}\Tk{J}\Tk{L}) q_i \\
b_\kappa \cdot b_{ij} \CC_{IJ}
&= \sum_{R,q_i , q_j} x_{R,q_i,q_j}  \tr_R ( \Tk{I} \Tk{J}) q_i q_j
\end{split}
\end{align}

As before, we can write all the elements of the right-hand sides as
a sum of products of the charge of each vector or hypermultiplet
under $\sT_{I,\kappa}$, $U(1)_i$ or $U(1)_j$.
Each charged hypermultiplet has a corresponding type I rational curve,
and its charge is given by the intersection numbers of the
curve with the four-cycles of type C or S.
Rewriting the right-hand sides of the equations we obtain
\begin{align}
\begin{split}
0 &= \sum_r (c_r \cdot T_{I,\kappa})(c_r \cdot T_{J,\kappa})(c_r \cdot T_{K,\kappa})(c_r \cdot S_i) \\
b_\kappa \cdot b_{ij} \CC_{IJ,\kappa}
&= \sum_r (c_r \cdot T_{I,\kappa})(c_r \cdot T_{J,\kappa})(c_r \cdot S_i)(c_r \cdot S_j)
\end{split}
\end{align}

Using
\be
b_\kappa \CC_{IJ,\kappa} = -\pi(\Tk{I} \cdot \Tk{J}),
\quad \pi(\Tk{I} \cdot S_i) =0,
\quad b_{ij}=-\pi(S_i \cdot S_j)\,,
\ee
we obtain the final expressions by rewriting the equations:
\begin{align}
\begin{split}
\pi (\Tk{I} \cdot S_i) \cdot \pi (\Tk{J} \cdot \Tk{L}) &+ \text{(2 other groupings)} \\
&= \sum_r (c_r \cdot T_{I,\kappa})(c_r \cdot T_{J,\kappa})(c_r \cdot T_{K,\kappa})(c_r \cdot S_i) \\
&+ \sum_\rho (2 g_\rho-2)(\chi_\rho \cdot T_{I,\kappa})(\chi_\rho \cdot T_{J,\kappa})(\chi_\rho \cdot T_{K,\kappa})(\chi_\rho \cdot S_i) \\
\pi (\Tk{I} \cdot \Tk{J}) \cdot \pi (S_i \cdot S_j) &+ \text{(2 other groupings)} \\
&= \sum_r (c_r \cdot T_{I,\kappa})(c_r \cdot T_{J,\kappa})(c_r \cdot S_i)(c_r \cdot S_j) \\
&+ \sum_\rho (2 g_\rho-2)(\chi_\rho \cdot T_{I,\kappa})(\chi_\rho \cdot T_{J,\kappa})(\chi_\rho \cdot S_i)(\chi_\rho \cdot S_j)
\end{split}
\end{align}
Note that the left hand side of the first equation is zero,
and that the two other groupings in the second equation are zero.
The second term on the right hand side of both equations are zero since
$S_i \cdot \chi_\rho =0$ by construction of type S cycles.
$\Box$

%--------------- Bibliography ---------------------------

\end{document}